\documentclass[aps,nofootinbib,twocolumn,superscriptaddress]{revtex4}

\usepackage[strict]{revquantum}

\newcommand{\figurefolder}{../fig}

\newcommand{\mps}{x}
\newcommand{\eps}{e}
\newcommand{\data}{d}

\newcommand{\MLE}{\text{MLE}}
\newcommand{\ESM}{\text{ESM}}

\newcommand{\calib}{{\text{cal}}}
\newcommand{\Rabi}{\text{Rabi}}
\newcommand{\Ramsey}{\text{Ramsey}}
\newcommand{\te}{t_\text{e}}
\newcommand{\tp}{t_\text{p}}
\newcommand{\tw}{t_\text{w}}
\newcommand{\tm}{t_\text{m}}

\renewcommand{\H}{H}    % hamiltonian
\renewcommand{\L}{L}    % lindblad dissipator
\renewcommand{\S}{S}    % superoperator
\newcommand{\uw}{{\mu\text{w}}}

\newcommand{\apxref}[1]{\hyperref[#1]{Appendix~\ref{#1}}}

\renewcommand{\figurefolder}{./fig}
\begin{document}

\title{Hamiltonian Learning with Online Bayesian Experiment Design in Practice}

\author{Ian Hincks}
\affilUWAMath
\affilIQC

\author{Thomas Alexander}
\affilUWPhys
\affilIQC

\author{Michal Kononenko}
\affilIQC
\affilUWChemEng

\author{Benjamin Soloway}
\affilIQC
\affilHaverford

\author{David G. Cory}
\affilUWChem
\affilIQC
\affilPI
\affilCIFAR

% new authors add your names

\date{\today}

\begin{abstract}
Estimating parameters of quantum systems is usually done by performing
a sequence of predetermined experiments and post-processing the resulting data.
It is known that online design, where the
choice of the next experiment is based on
the most up-to-date knowledge about the system,
can offer speedups to parameter estimation.
We apply online Bayesian experiment design to a Nitrogen
Vacancy (NV) in diamond to learn the values of a
five-parameter model describing its Hamiltonian and decoherence process.
Comparing this to standard pre-determined experiment sweeps,
we find that we can achieve median posterior variances on some parameters
that are between 10 and 100 times better given the same amount of data.
This has applications to NV magnetometry where one of the Hamiltonian
coefficients is the parameter of interest. Furthermore, the methods that we use are generic and can be adapted to any quantum device.
\end{abstract}

\maketitle

%=============================================================================
% MAIN DOCUMENT
%=============================================================================

%=============================================================================
\section{Introduction}
\label{sec:intro}
%=============================================================================

Characterizing quantum devices efficiently is an increasingly
important problem.
In the case of quantum processors, knowing system properties
and error processes is helpful for designing robust high-fidelity control.
If system parameters drift in time, they will
need to be periodically recharacterized, which reduces uptime.
Or, in the case of metrology, certain properties of the quantum
system are themselves the quantities of interest, and so more efficient
characterization leads to higher sensitivities.

Quantum system characterization is typically done by performing
a set of predetermined experiments and subsequently processing
statistics of the resulting data.
While there is nothing wrong with this---and indeed, in some cases,
this strategy can even be tuned to have near optimal performance---it
has long been known that \textit{online}
(also called \textit{adaptive} by some authors)
experiment design is generally
capable of outperforming predetermined experiment sweeps
\cite{chaloner_bayesian_1995,higgins_demonstrating_2009}.
As its name implies, online experiment design allows the next experiment
choice to depend somehow on what has already been learned.
The reason for the advantage is obvious---online
experiments can potentially avoid executing experiments that
are expected to be uninformative by using information that was
initially unavailable.

Online experiment design has a long history in quantum systems. Almost five decades ago, it was used to reduce the time required to
determine relaxation rates in NMR spin systems \cite{freeman_adaptive_1972},
and later to speed up inversion recovery $T_1$ measurements
\cite{taitelbaum_twostage_1993}.
In recent decades, it has been studied extensively, both in theory
and experiment, in the context of quantum phase estimation
\cite{
    wiseman_adaptive_1995,
    berry_optimal_2001,
    higgins_entanglementfree_2007,
    berry_how_2009,
    higgins_demonstrating_2009,
    xiang_entanglementenhanced_2011,
    yonezawa_quantumenhanced_2012,
    ciampini_quantumenhanced_2016}
and quantum state tomography
\cite{
    huszar_adaptive_2012,
    kravtsov_experimental_2013,
    ferrie_selfguided_2014,
    stenberg_adaptive_2015,
    struchalin_experimental_2016,
    granade_practical_2016,
    qi_adaptive_2017}.
Online experiment design has been suggested for sequence length choices in
randomized benchmarking experiments\cite{granade_accelerated_2015},
and adaptive protocols
to generate control pulses for quantum systems have been proposed
\cite{egger_adaptive_2014,ferrie_robust_2015,rol_restless_2017}.
Here we build on online experiment design applied to
quantum Hamiltonian estimation
\cite{
    sergeevich_characterization_2011,
    granade_robust_2012,
    ferrie_how_2013,
    wiebe_hamiltonian_2014,
    stenberg_simultaneous_2016,
    stenberg_characterization_2016},
where a Hamiltonian form (or set of forms) is specified, and
unknown coefficients of Hamiltonian terms are sought.

The purpose of the present work is to study
online Bayesian experiment design, with Hamiltonian estimation
as the inference problem of choice, using experimental data and noise on
a system with slightly non-trivial dynamics.
By non-trivial we mean that there are more than one or two
relevant inference parameters (we ultimately use 10, including nuisance
parameters describing optical drift), that
quantum state evolution does not admit a nice closed form
solution, and that we allow the ability to turn
on and off the control field within an experiment.
In doing so we hope to pave the way for similar experiments in
yet more complex systems.
To this end we interface a sequential Bayesian inference engine with
an experimental setup that controls the qutrit manifold of a single Nitrogen
Vacancy (NV) defect in diamond.
NV defects are widely studied quantum systems that can
be initialized and read-out optically
\cite{
    gruber_scanning_1997,
    jelezko_readout_2004,
    harrison_optical_2004},
manipulated at microwave frequencies\cite{jelezko_observation_2004},
and have long coherence times at room temperature\cite{balasubramanian_ultralong_2009}.
Their proposed applications include quantum sensing
\cite{
    dolde_electricfield_2011,
    acosta_temperature_2010,
    rondin_magnetometry_2014a}
and building quantum repeaters\cite{childress_faulttolerant_2006}.

This paper proceeds as follows.
In \autoref{sec:inference} we briefly overview statistical
inference, followed by a short summary of Bayesian experiment
design in \autoref{sec:experimental-design}.
In \autoref{sec:system-model} we define a system model for
the NV system in particular, whereas the previous sections
were general.
In \autoref{sec:computation-and-hardware} we discuss some
hardware, software, and implementation details of our setup.
Finally, in \autoref{sec:results} we present our results
comparing offline and online experiment design heuristics.
Code and data to reproduce the results of this paper can
be found in Reference~\cite{hincks_code_2018}.

%=============================================================================
\section{Inference of Quantum Devices}
\label{sec:inference}
%=============================================================================

We begin by defining some notation while reviewing parameter
estimation as applied to quantum devices.

Information about a quantum device can be encoded into a list of
real values, which we call \textit{model parameters}, labeled $\mps$.
For example, in the case of Hamiltonian learning,
these values parameterize the Hamiltonian operator of the
quantum system, or in the case of
state tomography, the entries of a density operator.
This set of parameters includes both parameters of interest, which
one is interested in learning, and nuisance parameters, which are
not of principle interest, but are still necessary to sufficiently
describe the system.

Quantum devices are controlled by some collection of
classical knobs that adjust various settings
such as power, timings, carrier frequencies, and so on.
We refer to a specific assignment of all of these settings as an
\textit{experiment configuration}, sometimes called the control
variables, which we label $\eps$.
Then an \textit{experiment}
consists of a quantum measurement (or set of quantum measurements)
made using this fixed experiment configuration.
For example, in this nomenclature, a standard Rabi curve
would be constructed by making a set of
experiments, each one defining---among other fixed parameters---a
pulsing time in its experimental configuration,
$\eps=(\ldots,t_\text{pulse},\ldots)$.

An experiment returns a datum $\data$.
This might be a photon count
over a known time interval, a time series of voltages,
or a number of `down' strong measurement results out of $N$
repetitions, and so on.

Generally, the goal of statistical inference is to learn the parameters
$\mps$ given a data set $\data_1,\ldots,\data_n$ with respective
configurations $\eps_1,\ldots,\eps_n$.
This requires us to additionally specify a model for the
system---something which connects the model parameters to the experiment
configurations and data.
This is done through a likelihood function,
\begin{equation}
    \Lhood(\mps;\data_{1:n},\eps_{1:n})
        = \Pr(\data_{1:n}|\mps,\eps_{1:n}),
    \label{eq:general-likelihood-function}
\end{equation}
which returns the probability of receiving a given dataset conditioned
on a hypothetical configuration $\mps$.
Here, and throughout this paper, we use subscripted index-range notation,
where, for example, $\data_{1:n}=\{\data_1,...,\data_n\}$.
Note that multiple models can be considered and compared---known
as model selection---if the
true model is not known.
For quantum systems, these likelihood models come naturally
through quantum system evolution formulas in conjunction
with Born's rule.

One popular inference choice is to maximize the likelihood function
with respect to $\mps$, producing the maximum likelihood estimate (MLE)
$\hat{\mps}_\MLE:=\operatorname{argmax}_\mps \Lhood$.
Confidence regions of this estimate can be constructed
with statistical derivations, or more generally, through techniques like bootstrapping.
Least-squared curve fitting is often used as a proxy for the MLE (with
confidence intervals arriving from assuming
a linearized model) since it is exactly equal to the MLE
for linear models and normal likelihood functions.

The MLE is one example of an estimator in a vast literature on estimator theory.
In the present work, we limit ourselves to the use of Bayesian inference because of its
natural integration with online experiments, discussed below.
In short, in the paradigm of (sequential) Bayesian inference, one maintains the most current state of knowledge about the model parameters $\mps$, encoded
as a probability distribution $\pi_n(\mps)=\Pr(\mps|\data_{1:n},\eps_{1:n})$,
where $n=1,2,3,...$ indexes the state of knowledge when the first $n$ data
points $\data_{1:n}$ have been collected and processed from the
first $n$ experiments $\eps_{1:n}$.
We write $\pi_0(x)$ to denote the distribution prior to all measurements.
The update from $\pi_{n-1}$ to $\pi_n$ is done through Bayes' law,
\begin{equation}
    \pi_n(\eps)
        = \frac{
            \Pr(\data_n|\mps,\eps_n)\pi_{n-1}(\mps)
        }{
            \Pr(\data_n|\eps_n)
        },
\end{equation}
so that our knowledge is improved sequentially as each datum arrives.
Note that the chain rule of conditional probabilities can be used to expand
this equation into
$\pi_n(\eps)=\Pr(\data_{1:n}|\mps,\eps_{1:n})\pi_0(\mps)/\Pr(\data_{1:n}|\eps_{1:n})$.

%=============================================================================
\section{Bayesian Experimental Design}
\label{sec:experimental-design}
%=============================================================================

\begin{figure*}
    \includegraphics[width=\textwidth]{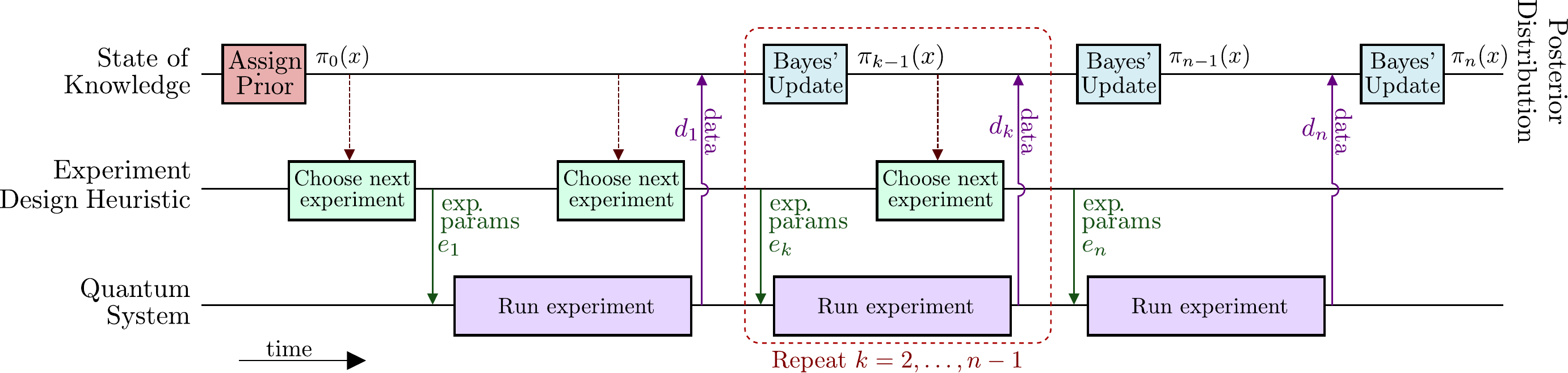}
    \caption{Timing diagram of online Bayesian learning. The role of the
    experiment design heuristic is to pick the next experiment configuration
    $e_{n+1}$, possibly based
    on the current state of knowledge, $\pi_n(\mps)$, resulting in the
    new data point $d_{n+1}$.
    This choice of experiment be computationally expensive, and is
    therefore run concurrently with quantum experiments.}
    \label{fig:online-timing-diagram}
\end{figure*}

An \textit{experiment design heuristic} is simply a function that
determines the next experiment configuration to use.
We say such a heuristic is \textit{online} if it explicitly uses the
results of preceding experiments, and we call it \textit{offline}
otherwise.
An experiment design timing diagram is shown in \autoref{fig:online-timing-diagram}.
Conventionally, as an example, Rabi curves are generated
with offline heuristics, where the next experiment is chosen by
increasing the pulse time by a fixed duration in each experiment.
The number of experiments and pulse time increments are usually
chosen through Nyquist considerations based on prior implicit
beliefs about the frequencies and relaxation times of the system.

We restrict our online design heuristics to Bayesian designs,
summarized in the following framework.
Let $U_n(\mps,\data,\eps)$ be the utility of collecting the datum
$\data$ under configuration $\eps$ given the hypothetical
model parameters $\mps$ and the current state of knowledge $\pi_n(\mps)$,
where a large value is good.
Using the Bayesian maxim of marginalizing over unknown quantities,
the average utility of observing $\data$ at step $n+1$ under
the possible experiment configuration $\eps$ is
\begin{align}
    U_n(\data,\eps)
        &= \int \tilde{\pi}_{n,\data,\eps}(\mps)U_n(\mps,\data,\eps)\dd\mps.
\label{eq:future-utility}
\end{align}
where
$\tilde{\pi}_{n,\data,\eps}(\mps)\propto \Lhood(\mps;\data,\eps)\pi_n(\mps)$
is the hypothetical posterior at step $n+1$ assuming $\data$ will be observed.
Since we do not know \textit{a priori} which $\data$ will
occur, the average utility of the possible configuration $\eps$ as a whole is
\begin{align}
    U_n(\eps)
        &= \int\Pr(\data|\eps) U_n(\data,\eps) \dd\data.
\label{eq:average-utility}
\end{align}
where $\Pr(\data|\eps)=\int \Pr(d|x,e)\pi_n(x)\dd x$ is the predictive
distribution\footnote{Note that
$
    \tilde{\pi}_{n,\data,\eps}(\mps)\Pr(\data|\eps)
        =\Pr(\mps,\data|\mps_{1:n},\data_{1:n},\eps),
$
and therefore $U_n(\eps)$ is the joint average over $\mps$ and $\data$ of
$U_n(\mps,\data,\eps)$ given the current state knowledge---this is a description
some may prefer to the two-step description involving
the intermediate quantity $U_n(\data,\eps)$ provided in the main-body.}.
Based on this quantity we can choose the next experiment to be
the one that maximizes the utility,
\begin{align}
    \eps_{n+1} = \argmax_\eps U_n(\eps),
\end{align}
with the maximum taken over some space of possible experiments.
If computed numerically, we might only hope to find local maxima.

One can consider different choices of utility function $U$.
When the application is inference of a non-linear system, such as
ours, it is common to choose a utility based on
mean-squared error \cite{chaloner_bayesian_1995}.
In particular, we choose $U_n=-r_{n,Q}$ where
\begin{align}
    r_{n,Q}(\mps,\data,\eps)
        &= \Tr\left[Q(\mps - \hat\mps_{n,\data,\eps})^\T (\mps - \hat\mps_{n,\data,\eps})\right]
\end{align}
where $Q$ is a positive semi-definite weighting matrix.
Here, $\hat{\mps}_{n,\data,\eps}=\int \mps \tilde{\pi}_{n,\data,\eps}(\mps)\dd\mps$
is the Bayes estimator of $\mps$.
In this case, $r_{n,Q}(\eps)$ has the simple interpretation of being the
expected posterior covariance matrix weighted against $Q$,
\begin{align}
    r_{n,Q}(\eps) = \Tr \left[Q\expect_\data[\Cov_{\tilde{\pi}}
        [\mps|\data,\eps]]\right],
    \label{eq:bayes-risk}
\end{align}
a quantity known as the $Q$-weighted
mean-squared-error \textit{Bayes risk} (some numerical implementation
details are outlined in \autoref{apx:brute-force-bayes-risk}).

%=============================================================================
\section{Nitrogen Vacancy System Model}
\label{sec:system-model}
%=============================================================================

The quantum system used in our experiment is a nitrogen vacancy (NV) center,
which is a defect found in diamond consisting of a nitrogen adjacent to
a vacant lattice position \cite{doherty_nitrogenvacancy_2013}.
Our goal for this section is to explicitly define model parameters, experiment
configurations, and a likelihood function for this system.
Once this is achieved, we will be able to employ sequential
Bayesian inference and online experiment design.

When in its stable negatively charged configuration, NV$^-$,
the vacancy is filled with
six electrons that form an effective spin-1 particle in the optical
ground state---this three level subspace comprises the system of interest.
The eigenstates are labeled $\ket{1}$, $\ket{0}$, and $\ket{-1}$, respectively
corresponding to the eigenvalues of the spin-1 operator $\Sz=\diag(1,0,-1)$.
There is a zero field splitting (ZFS) of $D\approx\SI{2.87}{GHz}$
between $\ket{0}$ and the $\operatorname{span}(\ket{-1},\ket{+1})$ manifold
that---at low fields ($\lesssim\SI{100}{G}$)---is the dominant energy term,
defining our $z$-axis.
The Zeeman splitting between the states $\ket{-1}$ and $\ket{+1}$
is determined by the magnetic field projection onto the $z$-axis, equal to
$\omega_e=\gamma_e |B_z|$ in the secular approximation,
where $\gamma_e\approx\SI{2.80}{MHz/G}$.
Spin manipulation is achieved with resonant microwave driving near
the transitions $D\pm\omega_e$.
Long coherence times are observed at room temperature, where
the spin state can be initialized and measured optically, and single
defects are studied in isolation using confocal microscopy.

In the rotating frame $\omega_\uw \Sz^2$,
with the rotating wave and secular approximations,
the Hamiltonian of the optical ground state is given by
\begin{align}
    \H/2\pi &= (D-\omega_\uw)\Sz^2 + (\omega_e+A \Iz)\Sz + \Omega_1(t)\Sx
\end{align}
where $(\Sx,\Sy,\Sz)$ are the spin-1 operators, $\omega_\uw$ is the applied
microwave frequency, $\Omega_1(t)$ is the microwave drive strength, $A$ is
the hyperfine splitting  due to the adjacent nitrogen-14 atom, and $\Iz$
is the nitrogen spin-1 operator along $z$.
Along with the $T_2^*$ decoherence time that introduces the
Lindblad operator $\L=\sqrt{1/T_2^*}\Sz$, these parameters are sufficient
to simulate the experiments that we perform.
Therefore, the model parameters of our spin system (a few
more nuisance parameters will be added later) are given by
\begin{align}
    \mps=(\Omega,\omega_e,\delta D,A,(T_2^*)^{-1})
\end{align}
where $\delta D=D-\SI{2.87}{GHz}$.
Here, $\Omega$ is the maximum possible
value that $\Omega_1(t)$ can take, so that we can write
$\Omega_1(t)=a(t)\Omega$ using the unitless pulse-profile function
$a(t):[0,\te]\rightarrow [-1,1]$ of duration $\te$.

\begin{figure}[t]
    \includegraphics[width=\textwidth]{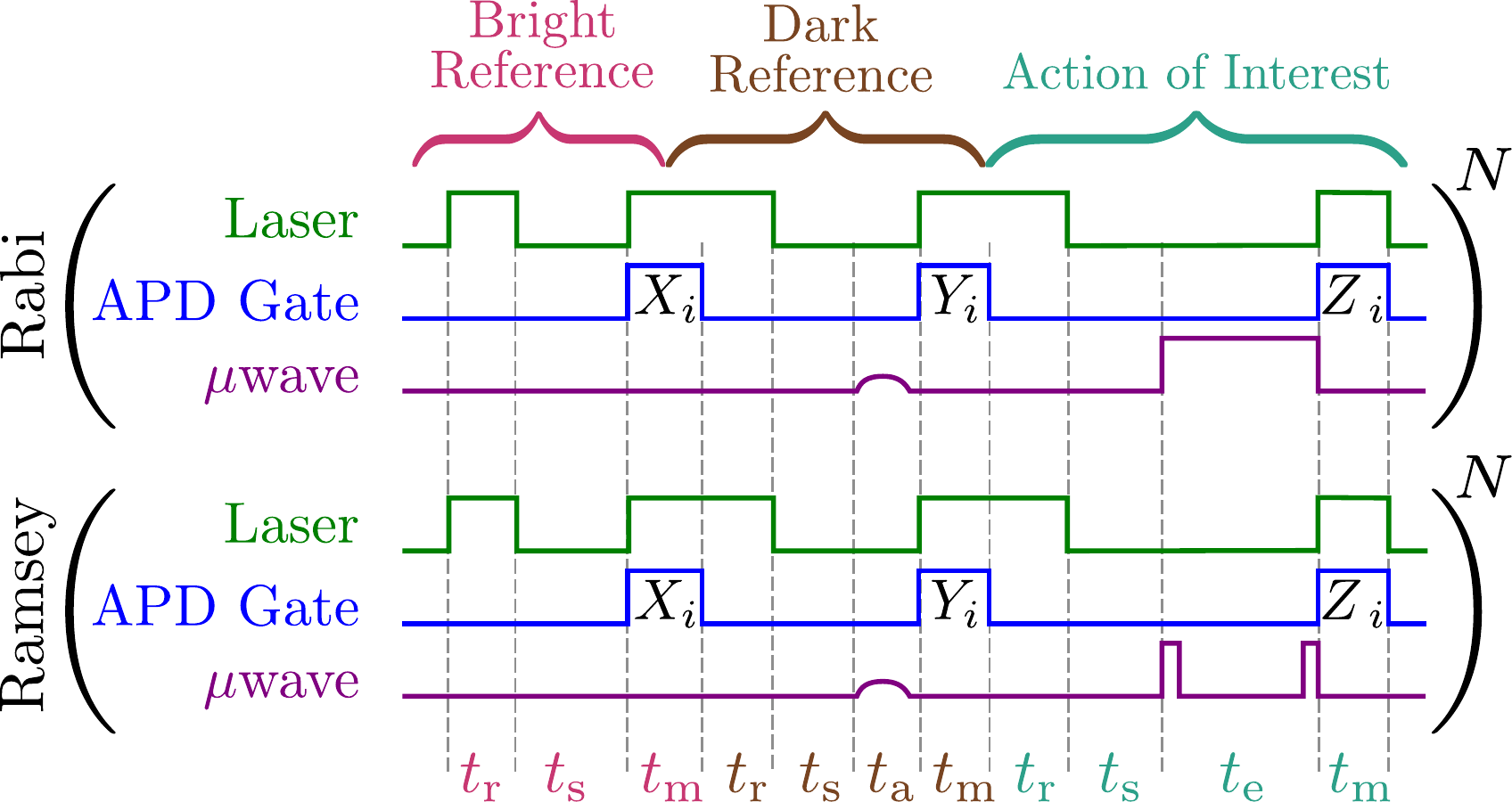}
    \caption{Pulse timing diagrams for Rabi (top) and Ramsey
        (bottom) experiments. An experiment has three control lines:
        whether the laser is on or off, whether the APD is counting
        photons or not, and the microwave amplitude profile.
        The pulse sequence is repeated $N$ times, collecting
        photon counts $(X_i,Y_i,Z_i)$ for $i=1,...,N$ for the bright reference,
        dark reference, and experiment, respectively, and finally summing them
        each over $i$ to produce the data point $\data=(X,Y,Z)$.
        Initial states are prepared by lasing for time $t_\text{r}$
        and letting the system settle for time $t_\text{s}$.
        Measurements consist of detecting photons for
        durations of length $\tm$ while lasing.
        The dark reference includes an adiabatic pulse of length
        $t_\text{a}$ which causes the state
        transfer $\ket{0}\rightarrow\ket{+1}$.
        The action of interest implements the microwave envelope
        $\Omega_1(t)$ of duration $\te$.
        Relative timing is not to scale in this diagram.
        }
    \label{fig:experiment-pulse-sequences}
\end{figure}

A general experiment configuration is then specified by
\begin{align}
    \eps=(a(t), \omega_\uw, N)
\end{align}
where $a(t)$ pulse profile, $\omega_\uw$ is the applied microwave frequency,
and $N$ is the number of repetitions of this experiment\footnote{The
experiment configuration must also specify values for each of the timings
labeled in \autoref{fig:experiment-pulse-sequences}, but as they are
calibrated independently from the experiment of interest, we omit
them here for simplicity.}.
In this paper, we restrict our attention to two special
cases of this general form, depicted in
\autoref{fig:experiment-pulse-sequences}, given by
\begin{enumerate}
    \item Rabi experiments, $\eps_\Rabi=(\tp,\omega_\uw, N)$,
    $a(t)=1$ for all $0\leq t\leq \te=\tp$; and
    \item Ramsey experiments, $\eps_\Ramsey=(\tp,\tw,\omega_\uw, N)$,
    \begin{equation*}
        a(t)=\begin{cases}
            0 & \tp<t<\tp+\tw\\
            1 & \text{else}
        \end{cases}
    \end{equation*}
    for all $0\leq t\leq \te=2\tp+\tw$.
\end{enumerate}

Given a hypothetical set of model parameters $\mps$ and
an experiment configuration $\eps$, the superoperator
(in column-stacking convention) is given by the solution
to the Lindlad master equation,
\begin{subequations}
\begin{align}
    S(\mps,\eps)
        &=\mathcal{T}\e^{\int_0^{\te}(C[\H(t)]+D[\L])\dd t},
        \quad\text{ where } \\
    C[\H(t)]
        &=-i(\I\otimes\H(t)-\overline{\H(t)}\otimes\I)
        \quad\text{ and } \\
    D[L]
        &= \overline{L}\otimes L
            -(\I\otimes L^\dagger L + \overline{L^\dagger L}\otimes \I) / 2,
\end{align}
\end{subequations}
and where $\mathcal{T}$ is Dyson's time ordering operator.
This results in the measurement probability
\begin{align}
    p(\mps,\eps)
        &=  \dbra{P_0} \S(\mps,\eps)\dket{\rho_0},
    \label{eq:born-rule}
\end{align}
where our initial state is $\rho_0=\ketbra{0}\otimes\I/3$
and the measurement projector is $P_0=3\rho_0$.

The standard measurement protocol of the NV system at room
temperature does not have direct access to strong measurements
\cite{hincks_statistical_2018}.
Instead, the probability $p(\mps,\eps)$ is obstructed by three Poisson
rates, so that data is in the form of a triple $\data=(X,Y,Z)$
where
\begin{subequations}
\begin{align}
    X|\alpha
        & \sim\poissondist(N\alpha) \\
    Y|\beta
        & \sim\poissondist(N\beta) \\
    Z|\mps,\eps,\alpha,\beta
        & \sim\poissondist(N(\beta+p(\mps,\eps)(\alpha-\beta)))
\end{align}
\label{eq:nv-measurement}%
\end{subequations}
with $\alpha$ and $\beta$,
the number of expected
photons for the bright and dark references in a single shot with a
given measurement duration $\tm$,
 satisfying $0<\beta<\alpha$.
The values $\alpha$ and $\beta$ are nuisance parameters which we
must append to our model parameters, giving
\begin{align}
    \mps=(\Omega,\omega_e,\delta D,A,(T_2^*)^{-1},\alpha,\beta).
    \label{eq:nv-model-parameters}
\end{align}

\begin{figure}
    \includegraphics[width=\textwidth]{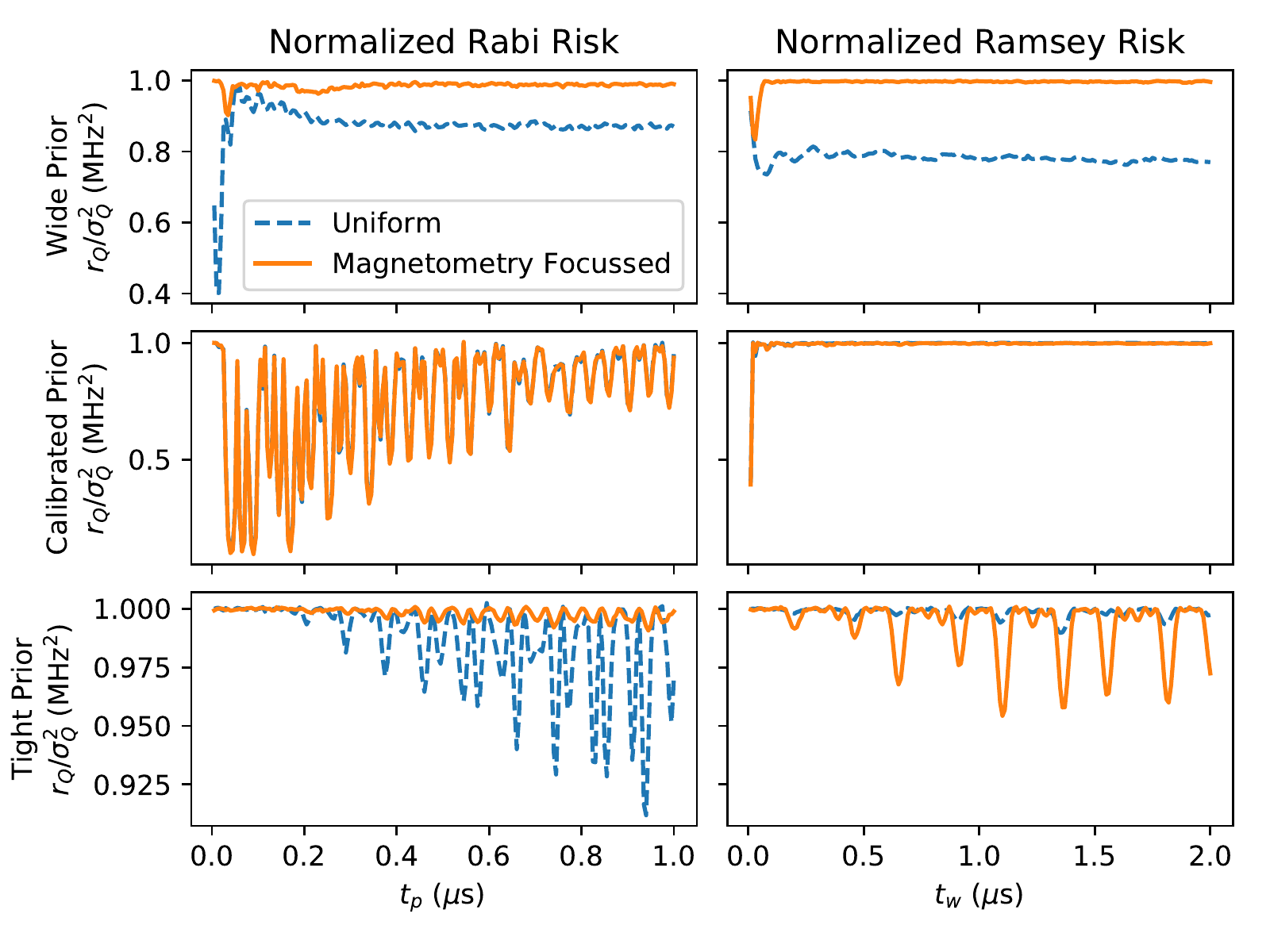}
    \caption{Calculation of risk for three different prior distributions (rows)
        and for both Rabi and Ramsey type experiments (columns).
        The dashed blue lines use a uniform weight matrix $Q=\diag(1,1,1,1,1)$,
        and the solid orange lines use a weight matrix focused only on $\omega_e$,
        $Q=\diag(0,1,0,0,0)$.
        Values have been normalized against $\sigma_Q^2=\Tr(Q\Cov_{\pi}[\mps])$
        where $\Cov_{\pi}[\mps]$ is the covariance matrix of a prior
        distribution $\pi$,
        so that, for example, a value of $r_Q(\eps)/\sigma_Q^2=0.95$ for a given
        experiment $\eps$
        implies a $5\%$ expected improvement in weighted covariance.
        The wide prior (top row) is defined in \autoref{eq:wide-prior}, the
        calibrated prior (middle row) is defined in \autoref{eq:calibrated-prior},
        and the tight prior (bottom row) is the same as the calibrated prior, but
        without widening the $\omega_e$ parameter.
        Note that the Rabi and Ramsey experiments share a $y$-axis on each row.
        We see that, among these examples, the only beneficial setting to
        perform a Ramsey experiment is with the tight prior when $\omega_e$
        is the parameter of interest.
        }
    \label{fig:risk-summary}
\end{figure}

The likelihood function (see \autoref{eq:general-likelihood-function})
for a single experiment is then given by
\begin{align}
    \Lhood(\mps;\data,\eps)
        &= f(X,N\alpha)
            \cdot f(Y, N\beta) \nonumber \\
            &\quad\quad\times f(Z, N(\beta+p(\mps,\eps)(\alpha-\beta))
    \label{eq:likelihood-function}
\end{align}
where $f$ is the probability mass function of the Poisson distribution,
$f(Q,\lambda) = \e^{-\lambda}\lambda^Q/Q!$.
Some example risk plots (\autoref{eq:bayes-risk}) of this model
are shown in \autoref{fig:risk-summary}.

%=============================================================================
\section{Computation and Hardware}
\label{sec:computation-and-hardware}
%=============================================================================

For all experiment design heuristics, offline and online,
we use the sequential Monte Carlo (SMC) \cite{doucet_tutorial_2009}
method to numerically compute sequential posteriors
using the Python library \qinfer  \cite{granade_qinfer_2017}.
In this algorithm, the state of knowledge about the model
parameters, $\pi_n(\mps)$,
is approximated as a finite list of weighted hypothetical values
(which are called \textit{particles}),
\begin{align}
    \pi_n(\mps)
        &= \sum_{i=1}^{K}w_{n,i} \delta(\mps-\mps_{n,i}),
    \label{eq:particle-approximation}
\end{align}
where $w_{n,i}\geq 0$ with $\sum_{i=1}^K w_{n,i}=1$, and where
$\delta(\cdot)$ is the delta mass distribution centered at $0$.
The particle-approximated prior, $\pi_0(\mps)$, is generated by
sampling $K$ initial particles $\mps_{0,i}$ from the prior distribution
and setting uniform weights $w_{0,i}=1/K$.
Given the new datum $\data_{n+1}$ under experiment
configuration $\eps_{n+1}$,
Bayes update can be implemented with the simple multiplication
\begin{align}
    w_{n+1,i}
        &\propto w_{n,i}\cdot \Lhood(\mps_{n,i};\data_{n+1},\eps_{n+1})
    \label{eq:particle-update}
\end{align}
which requires $K$ simulations of the quantum system to
compute the likelihoods (\autoref{eq:likelihood-function}), and where
the constant of proportionality is chosen so that
$\sum_{i=1}^K w_{n+1,i}=1$.
We use the scheme of Liu and West \cite{liu_combined_2001}
to resample particle locations,
triggered by a threshold in the effective particle count,
$n_\text{eff}:=1/\sum_{i=1} w_{n,i}^2$ \cite{granade_qinfer_2017}.
We also use the bridged-updating trick discussed in
Reference~\cite{hincks_statistical_2018}.

We note that the expensive stage of this algorithm is
embarrassingly parallel---simulations under the various
model parameters $\mps_{n,i}$ can be performed independently.
All of our processing was run on a desktop computer with simulations
parallelized over the 12 cores on a pair of Intel Xeon X5675 CPUs.
In this configuration, our updates took on the order of $2$ seconds
with $K=30000$ particles.
In principle, simulations could instead be run on quantum simulators,
as was recently demonstrated \cite{wang_experimental_2017}.

For online heuristics, the Bayes risk (\autoref{eq:bayes-risk}) is calculated by
noting that the particle approximation turns all integrals, which includes
expectations and covariances, into finite sums---see
\autoref{apx:brute-force-bayes-risk} for details.
Some risk calculations for the NV model
are plotted in \autoref{fig:risk-summary}.
As seen in the timing diagram in \autoref{fig:online-timing-diagram}, these
calculations (along with the Bayes updates) are
performed concurrently with experiments so that they do not
add to experiment cost\footnote{Of course, this is only
possible so long as the experiment repetition count is large
enough compared to the parallelized simulation cost. In our setup, at
our count rates, we landed naturally in this regime with
CPU computation a single desktop computer.}.
This causes the side-effect where the next experiment is selected using
information that is one cycle out-of-date; however,
in our simulations at our data collection rates,
we found that this did not have a noticeable
effect on learning rates.
A new experiment configuration $\eps_{n+1}$ having been
decided, by whatever heuristic,
the processing computer sends $\eps_{n+1}$ to the
computer which controls experiments.
The experiment is run, and the datum $\data_{n+1}=(X,Y,Z)$
is returned to the processing computer.
This process is iterated until some stopping criterion is met---for
example, in our experiments, we chose to
stop after 200 experiments had been performed.

In our setup, the processing computer and the experiment
computer communicate over ethernet with TCP.
We use a custom built confocal microscope to isolate an individual
NV center in bulk diamond.
All of our experiments were performed on the same NV center.
Laser light is produced by a continuous-wave \SI{100}{mW} laser
at \SI{532}{nm}, and switched using a double pass through an
acousto-optic modulator. %(Isomet 1250C-84).
Photons are collected with an avalanche-photo detector
(APD). %; Excelitas SPCM-AQRH)
%and counted with a National Instruments data acquisition card.
Microwaves are transmitted to the NV by an antenna of
diameter \SI{25}{um} about \SI{100}{um} away from the defect,
generated by a microwave synthesizer, and shaped by two channels of
an arbitrary waveform generator (AWG) %(Tektronix AWG 5002)
that mix via an IQ modulation. %(Marki IQ0255LMP).
Experimental configurations are manifest as waveforms on the AWG.
We use a caching strategy, where the experiment computer
uses a hash table to check if the desired experiment already
exists in the AWGs memory, avoiding data transfer costs when possible.

%=============================================================================
\section{Effective strong measurements and drift tracking}
\label{sec:esm-and-drift-tracking}
%=============================================================================

\begin{figure}
    \centering
    \includegraphics[width=\textwidth]{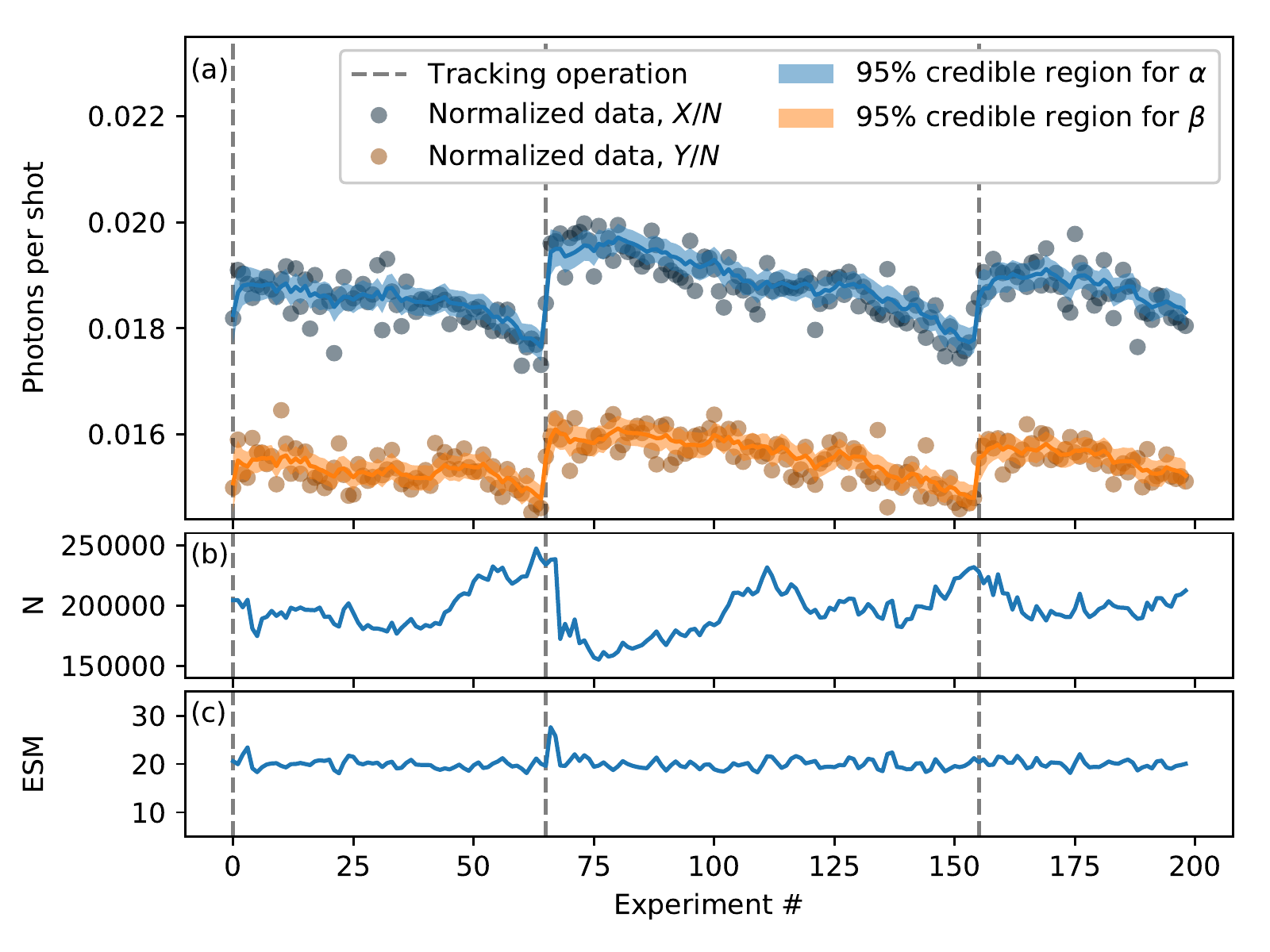}
    \caption{An NV drift tracking example, where tracking
    operations take place at the vertical dashed lines.
    (a) Sub-poissonian $95\%$ credible regions are shown on top of data
    normalized by the experiment repetition count, $N$.
    (b) The repetition count was chosen online to maintain a
    constant ESM value of $20$, which is plotted in (c).
    Several hundred trials were searched through
    to find this extreme but illustrative example---references are typically
    quite flat.}
    \label{fig:tracking-example}
\end{figure}

The amount of information provided by a measurement of $Z$ (see
\autoref{eq:nv-measurement}) depends on the values
of $\alpha$ and $\beta$.
Their magnitudes, relative contrast, and uncertainty
all contribute to this information content.
We quantify this idea by introducing what we call
the number of \textit{effective strong measurements} (ESM),
defined as the number of two-outcome strong measurements one would
(hypothetically) have to do to gain the equivalent amount of
information about $p(\mps,\eps)$, averaged uniformly over $p\in[0,1]$.
This works out to
\begin{align}
    \ESM = \frac{
            (\hat\alpha-\hat\beta)^2
        }{
            3(\hat\alpha+\hat\beta)+2\left(\sigma_\alpha^2+\sigma_\beta^2\right)
        }.
\end{align}
where $\hat\alpha$ and $\hat\beta$ are our current estimates of
$\alpha$ and $\beta$, and $\sigma_\alpha$ and $\sigma_\beta$
are standard deviation uncertainties in these estimates.
See \autoref{apx:effective-strong-measurements} for details.
We choose the number of repetitions in the next experiment, $N$,
such that the expected value of $\ESM$ is constant---see
\subref{fig:tracking-example}{(b-c)}.
This is especially important for the purpose of our paper,
which is to compare experiment design heuristics.
In this way,
certain heuristics are not artificially improved because
of favorable lab conditions on a certain day of the week.

The true specific values of the references $\alpha$ and $\beta$ depend
not only on the optical dynamics of the quantum system itself,
but also on the quality of the microscope's alignment.
As the temperature of the lab changes, for instance, one
can expect the values of $\alpha$ and $\beta$ to drift
as the location of the NV center moves with respect to the
focal spot of the microscope.
To account for this, a tracking operation is
performed periodically, where the focus of microscope is
repositioned based on a new set of images taken with the microscope.

A model that assumes these reference values are constant
in time can lead to inaccurate results, or even failure.
To account for this drift,
we append a Gaussian random
walk model for the parameters $\alpha$ and $\beta$ to the static
model defined in \autoref{sec:system-model}.
Specifically, we assume that immediately prior to
a particle update (\autoref{eq:particle-update}) the reference
indices of the each model parameter particle
undergo a resampling step defined as
\begin{align}
    \matrixtwobyone{\alpha_{n,i}}{\beta_{n_i}}
        &\sim \normaldist\left(
            \matrixtwobyone{\alpha_{n,i}}{\beta_{n_i}},
            \Delta t
            \matrixtwobytwo{\sigma_\alpha^2}{\sigma_{\alpha,\beta}}{\sigma_{\alpha,\beta}}{\sigma_\beta^2}
            \right),
\end{align}
where $\Delta t$ is the amount of time elapsed since the last
update.
The hyper-parameters $\sigma_\alpha$, $\sigma_\beta$, and
$\sigma_{\alpha,\beta}$ are treated as unknown; they are appended to
the model parameters, and co-learned along with the parameters defined
in \autoref{eq:nv-model-parameters}.
We use a wide inverse Wishart distribution as the prior with a
degrees-of-freedom parameter $\nu=30$ and a scale matrix $\Psi$
such that the mean value of the prior corresponds to
$\sigma_\alpha=\sigma_\beta=\sigma_{\alpha,\beta}/0.7=\SI{0.036}{/hour}$.
We use an empirical prior on $\alpha$ and $\beta$, where before
the actual experiments take place, a reference-only experiment is performed
with $N=300000$ repetitions, and the prior
is set as $\alpha\sim\gammadist(\mu=X/N,\sigma=3\sqrt{X}/N)$
and $\beta\sim\gammadist(\mu=Y/N,\sigma=3\sqrt{Y}/N)$.
When a tracking operation is performed, the distribution of
$\alpha$ and $\beta$ is resampled from the prior $\pi_0(\mps)$,
with all other parameters of the model held fixed.
We chose to perform tracking operation at the start of each trial,
and each time our estimate of $\alpha$ dipped below
our prior estimate of $\alpha$ minus five times the standard
deviation of our prior for $\alpha$.

%=============================================================================
\section{Results}
\label{sec:results}
%=============================================================================

\newcommand{\mystrut}{\rule{0pt}{1.5\normalbaselineskip}}
\begin{table*}[t]
    \centering
    \begin{tabularx}{\textwidth}{m{15em}X}
        \textbf{Heuristic} & \textbf{Definition} \\
        \hline\mystrut
        Alternating Linear
            & Offline; Sequential alternation between elements of
                the experiment \\
                & sets
                $E_\Rabi(\SI{500}{ns},100)$ and
                $E_\Ramsey(\hat{t}_{\text{p,best}},\SI{2}{us},100)$ \\
        \hline\mystrut
        Ramsey Sweeps
            & Offline;
                Two back-to-back sweeps through
                the experiment set
                $E_\Ramsey(\hat{t}_{\text{p,best}},\SI{2}{us},100)$ \\
        \hline\mystrut
        Uniformly Weighted Risk
            & Online;
                $e_{n+1}=\underset{e\in E}{\argmax}\left(r_{n,Q}(e)\right)$
                where $Q=\diag(1,1,1,1,1)$ and \\
                & $E=E_\Rabi(\SI{500}{ns},100)\cup
                E_\Ramsey(\hat{t}_{\text{p,best}},\SI{2}{us},100)$ \\
        \hline\mystrut
        Magnetometry Weighted Risk
            & Online;
                    $e_{n+1}=\underset{e\in E}{\argmax}\left(r_{n,Q}(e)\right)$
                    where $Q=\diag(0,1,0,0,0)$ and \\
                    & $E=E_\Rabi(\SI{500}{ns},100)\cup
                    E_\Ramsey(\hat{t}_{\text{p,best}},\SI{2}{us},100)$ \\
        \hline
    \end{tabularx}
    \caption{Summary of heuristics used to choose experiments.
        The best Ramsey tip time is defined by
        $\hat{t}_{\text{p,best}}=1/(4\hat{\Omega})$ (rounded to the nearest
        \SI{2}{ns}), where $\hat\Omega$ is the current
        Bayes estimate of the microwave drive amplitude.
        $E_\Rabi(t_{\max},m)$ denotes a set of Rabi experiments with pulse
        times $\tp=t_{\max}/m,2t_{\max}/m,\ldots,t_{\max}$, and
        $E_\Ramsey(\tp, t_{\max},m)$ denotes a set of Ramsey experiments with wait times
        $\tw=t_{\max}/m,2t_{\max}/m,\ldots,t_{\max}$ and pulse times $\tp$.
        The components of weight matrices $Q$ correspond to the
        Hamiltonian parameters
        $(\Omega,\omega_e,\delta D,A,(T_2^*)^{-1})$, with
        zeros for reference parameters.
    }
    \label{tab:heuristics}
\end{table*}

There are many choices to be made, even for this small system.
For example, we have already limited ourselves to Rabi and
Ramsey experiments.
Put differently, and given that our free evolution commutes with
both our initial state and measurement,
we have limited ourselves to bang-bang control
with a maximum of two pulses.
This is to ease simulations (bang-bang), and to reduce the
search space for online heuristics (two pulses or fewer).
We simplify the situation further by choosing to work in the low field
regime, say $\lesssim\SI{3}{G}$.
This saves us from having to adaptively modify the synthesizer
frequency $\omega_\uw$; we keep a fixed value of
$\omega_\uw=\SI{2.87}{GHz}$ for all experiments.
It also
prevents us from having to make decisions about the relative
phase between the two Ramsey pulses, to which we are
almost entirely insensitive at low field and with linearly
polarized microwaves.
These particular choices are by no means necessary, but serve
as a starting place to explore the landscape.
From the perspective of metrology, these choices amount to studying
the efficiency of DC magnetometry at low field with the NV system using
the double quantum manifold.

In our first comparison between experiment design heuristics,
we use a wide prior on the Hamiltonian
parameters given by
\begin{subequations}
\begin{align}
    \Omega/\si{MHz}
        &\sim \uniformdist\left( [0, 20] \right), \\
    \omega_e/\si{MHz}
        &\sim \uniformdist\left( [0, 10] \right), \\
    \delta D/\si{MHz}
        &\sim \uniformdist\left( [-5, 5] \right), \\
    A/\si{MHz}
        &\sim \uniformdist\left( [1.5, 3.5] \right), \\
    T_2^*/\si{\micro\second}
        &\sim \uniformdist\left( [1, 20] \right).
\end{align}
\label{eq:wide-prior}%
\end{subequations}
along with the reference priors discussed in
\autoref{sec:esm-and-drift-tracking}.
We implement the offline heuristic \textit{Alternating Linear} and the
online heuristics \textit{Uniformly Weighted Risk} and
\textit{Magnetometry Weighted Risk}
defined in \autoref{tab:heuristics}.
The offline heuristic is motivated by standard DC magnetometry,
where, intuitively, Rabi experiments are used to determine the pulse length
that causes $\ket{0}\mapsto\frac{\ket{+1}+\ket{-1}}{\sqrt{2}}$,
and Ramsey experiments subsequently exploit  this superposition state
to measure the relative
phase accumulation between $\ket{+1}$ and $\ket{-1}$, which is
proportional to $\tw \omega_e$.
Note that, unconventionally, this heuristic alternates between
Rabi and Ramsey experiments, as was done in \cite{hincks_statistical_2018}---this improves
numerical stability of the SMC sampler; as different
experiments are statistically independent, alternation
does not affect the overall information content.
The two online experiments differ only in the weighting matrix $Q$ that is used---the
first weights all quantum system parameters equally, and the second projects risk
onto only one parameter, $\omega_e$.

\begin{figure*}[t!]
    \centering
    \includegraphics[width=\textwidth]{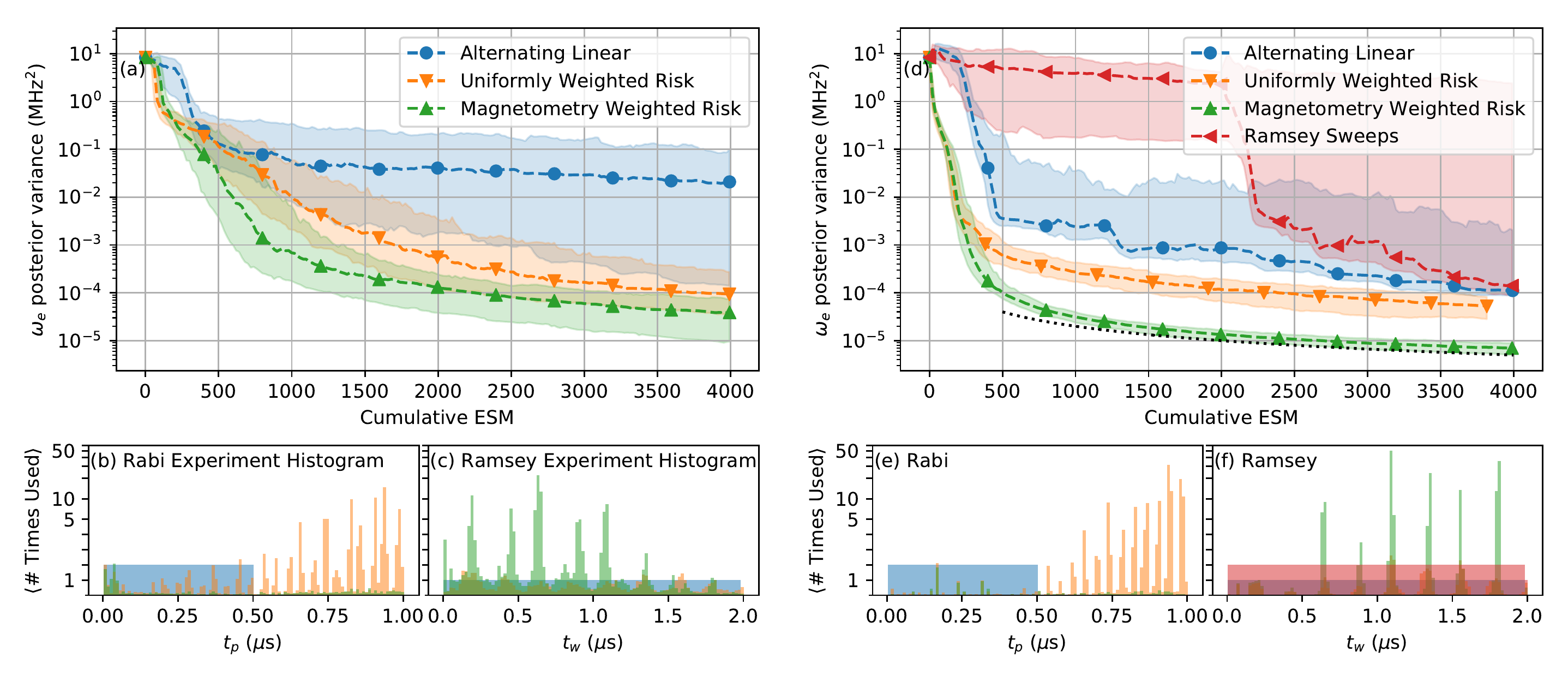}
    \caption{Comparison of experiment design heuristics
    (see \autoref{tab:heuristics}) where each heuristic was run
    with $100$ independent trials using $200$ experiments per trial.
    The left figures (a-c) use the wide prior of \autoref{eq:wide-prior},
    and the right figures (d-f) use the calibrated prior of
    \autoref{eq:calibrated-prior}.
    (a,d) For the parameter $\omega_e$, the median posterior variance
    over 100 trials is plotted (dashed lines), and
    regions between the $10\%$ and $90\%$ percentiles are shaded.
    The $x$-axes display ESM (effective strong measurements), where
    roughly $20$ effective bits of data are collected per experiment,
    see \autoref{sec:esm-and-drift-tracking}.
    The black dotted line scales as ESM$^{-1}$.
    In (b-c,e-f), histograms of which experiments each heuristic uses are
    shown, normalized to represent the average number of times
    used per trial.
    Note that the $y$-axis between histograms is shared, that
    the scaling switches from linear to logarithmic at $y=5$, and
    that all four subfigures contain $100$ histogram bins.
    Additional learning curves are plotted in \autoref{apx:supp-plots-and-data}.
    }
    \label{fig:heuristic-comparison}
\end{figure*}

Results of this first comparison
are shown in \subref{fig:heuristic-comparison}{(a-c)}.
Here, it is seen that both online heuristics outperform the offline heuristic, with a final
gap of a bit more than two orders of magnitude in the median (over trials)
posterior variance of $\omega_e$ after $4000$ ESM.
In the histograms we see that the magnetometry focused online heuristic
uses almost all Ramsey experiments, and the uniformly weighted online heuristics uses
almost all Rabi experiments, which agrees with the risk profiles
plotted in \autoref{fig:risk-summary}.
We see also that the offline heuristic has a much larger spread in posterior
variances across trials (area of shaded regions),
where some trials perform almost as well as the online heuristics,
but many perform significantly worse.
In this sense, in addition to tighter posteriors on average,
these online heuristics have the extra advantage of being
more reliable.
Our guess is that offline heuristics require luckily informative data at certain key
experiments to perform well, whereas online experiments can simply repeat these key
experiments.
Finally, note that the magnetometry focused
online heuristic slightly
outperforms the evenly weighted online heuristic---this
is unsurprising as we happen to be plotting the variance of the magnetometry
parameter, $\omega_e$.

In the context of magnetometry, it is unrealistic to assume such a
wide prior as given in \autoref{eq:wide-prior}.
More likely, one has already calibrated the quantum device and wants to
learn only the value of $\omega_e$.
For example, one might be constructing a magnetic image
\cite{
    maletinsky_robust_2012,
    grinolds_nanoscale_2013,
    rondin_strayfield_2013},
and each
pixel of the image requires a new field measurement.
Therefore, for our second comparison, we place a prior that is tight in all
Hamiltonian parameters except $\omega_e$, given as
\begin{subequations}
\begin{align}
    \omega_e/\si{MHz}
        &\sim \uniformdist\left( [0, 10] \right), \\
    (\Omega,\delta D,A,(T_2^*)^{-1})/\si{MHz}
        &\sim \normaldist\left(
            \mu_\calib,
            \Sigma_\calib
        \right),
\end{align}
\label{eq:calibrated-prior}%
\end{subequations}
where $\mu_\calib$ and $\Sigma_\calib$ are taken

from the posterior of a set of previously run calibration
experiments, see \autoref{apx:supp-plots-and-data} for
details.
In our study of this second prior, in addition to
the three heuristics used above,
we consider another heuristic called \textit{Ramsey Sweeps} that uses only
Ramsey experiments, since they are the \textit{de facto} method for
measuring static Hamiltonian terms along $z$.
Results for this second prior
are shown in \subref{fig:heuristic-comparison}{(d-f)}.
There are a few interesting features.
The first is that it is clearly visible where the \textit{Ramsey Sweeps}
heuristic finishes one sweep and starts the next, at 2000 ESM.
The second is that all three of the heuristics that were also used
for the wide prior (\autoref{eq:wide-prior}) have significantly
less spread under the calibrated prior.
The third is that the magnetometry weighted online heuristic has a much
clearer advantage over the uniformly weighted online heuristic than in
the case of the wide prior comparison.
Finally, notice in the histograms, that the uniformly weighted
online heuristic again chooses Rabi experiments almost
exclusively.

Supplementary plots, including posterior distributions, can be
found in \autoref{apx:supp-plots-and-data}.

Our online
learning rates appear to be at the standard quantum limit (SQL)
once transient
behavior has settled down; the dotted line in
\subref{fig:heuristic-comparison}{(d)} guides the eye with a
curve $\propto\text{ESM}^{-1}$.
The transient behavior prior to the SQL regime
looks qualitatively exponential as a function of ESM.
This does not violate the Heisenberg limit
($\sigma^2\propto \text{ESM}^{-2}$) because experiment times,
$\te$, are able to exponentially increase, too
 \cite{sergeevich_characterization_2011}.
Exponential-into-SQL scaling is consistent with previous
Hamiltonian estimation research, where the coherence time of the system
controls the transition location---ideally we would
perform Ramsey experiments with arbitrarily long wait times, but
finite $T_2^*$ makes such experiments uninformative
\cite{ferrie_how_2013}.

%=============================================================================
\section{Conclusions}
\label{sec:conclusions}
%=============================================================================

We compared the ability of several experiment design heuristics to
experimentally learn the electronic ground state Hamiltonian of an
NV defect in diamond.
Some of our heuristics were offline---using experiment sweeps that were
predetermined, and some of heuristics were online---using knowledge gained from
previous experiments to choose the next experiment adaptively.
The heuristics we used are summarized in \autoref{tab:heuristics}.
All data analysis was done with sequential Bayesian inference,
and all online heuristics were based on minimizing the
weighted Bayes risk over a collection of possible experiments.
Heuristics were compared by running 100 independent trials of each, and comparing
the reduction in posterior variance of certain parameters as a function of
the number of experiments performed.

We found that our online heuristics
outperformed our offline heuristics; results are summarized in
\autoref{fig:risk-summary}.
In particular, in the case of a very wide prior on all parameters
(\subref{fig:heuristic-comparison}{(a-c)}), we found that the
median posterior variance of the parameter $\omega_e$---which is proportional
to the external magnetic field's projection onto the z-axis---is over two
orders of magnitude smaller after 200 experiments (comprising ~200 effective
strong measurements per experiment) for the online heuristic called
\emph{Magnetometry Weighted Risk} than
it is for the offline heuristic called \emph{Alternating Linear}.
Next, in the case of a prior that is tight on all paramaters except $\omega_e$
(\subref{fig:heuristic-comparison}{(d-f)}), we found about an order of
magnitude of improvement between the best online heuristic and the
best offline heuristic.
The use case of this prior is when one wants to use a calibrated NV
device to measure many magnetic fields.

Consistent with intuition, we found that when online experiments are weighted
to improve $\omega_e$ alone, they tend to choose Ramsey experiments almost
exclusively, rather than Rabi experiments, see
\subref{fig:heuristic-comparison}{(b-c,e-f)}.

In addition to faster decrease in variance, we also found that variance
decreases more predictibly for online heuristics than it does for predetermined
heuristics.
This is seen in the tighter 80\% percentile regions of
\subref{fig:heuristic-comparison}{(a,d)} for online experiments.
For example, the difference in the final posterior variance of the parameter
$\omega_e$ varies by as much as four orders of magnitude between
independent trials for the \emph{Ramsey Sweeps} heuristic, whereas it
always varies by less than one order of magnitude for all online heuristics.

Studies of the sort presented here necessarily suffer from having to
make choices---in the end we had to choose a
small number of heuristics to compare, which types of experiments heuristics
should be allowed to perform, what the hyper-parameters of each heuristic
should be, what the initial prior over parameters should be, and so on.
Though these choices are ultimately arbitrary, we attempted to make them
reasonable, with the end goal of comparing a fully brute-force
Bayesian scheme against what have historically been the \emph{de facto}
methods of characterization.
While we would not be surprised to find a less computationally expensive
experiment design heuristic for this particular problem
with similar performance (for example, see
the heuristic policies in \cite{stenberg_characterization_2016}),
the advantage of a full-risk based approach is that it doesn't require an
expert to design a heuristic for every particular combination system and
protocol.
Indeed, minimizing Bayes risk, if computationally feasible either with
classical or quantum resources, is a sensible approach for practically
any characterization protocol, from tomography to randomized benchmarking.

%=============================================================================
% END MATTER
%=============================================================================

\acknowledgments{
The authors gratefully acknowledge contributions from the Canada First
Research Excellence Fund, Industry Canada, Canadian Excellence
Research Chairs, the Natural Sciences and Engineering
Research Council of Canada, the Canadian Institute for Advanced
Research, and the Province of Ontario.
}

\bibliographystyle{apsrev4-1}
\bibliography{nv-adaptive}

%=============================================================================
% APPENDICES
%=============================================================================

\appendix
\onecolumngrid

%=============================================================================
\section{Supplementary Plots and Data}
\label{apx:supp-plots-and-data}
%=============================================================================

The calibration prior of \autoref{eq:calibrated-prior} was generated
by processing two trials of the Alternating Linear heuristic, for
a total of 400 experiments, and roughly 8000 ESM.
The first two moments of the posterior distribution were computed, resulting in
the values
\begin{subequations}
\begin{align}
    \mu_\calib&=\begin{pmatrix}
        11.55 \\ -0.86 \\ 2.18 \\ 0.35
    \end{pmatrix}
    \si{MHz} \\
    \Sigma_\calib&=\begin{pmatrix}
        \num{2.56e-05} & \num{1.02e-03} & \num{7.67e-07} & \num{3.80e-05} \\
        \num{1.02e-03} & \num{1.06e-01} & \num{1.97e-04} & \num{2.50e-03} \\
        \num{7.67e-07} & \num{1.97e-04} & \num{7.51e-05} & \num{-1.02e-04} \\
        \num{3.80e-05} & \num{2.50e-03} & \num{-1.02e-04} & \num{1.01e-03}
    \end{pmatrix}
    \si{MHz^2}
\end{align}
\end{subequations}
for the ordered parameters $(\Omega,\delta D,A,(T_2^*)^{-1})$,
where $\omega_e$ and nuissance parameters ham been marginalized over.

\begin{figure*}
    \includegraphics[width=\textwidth]{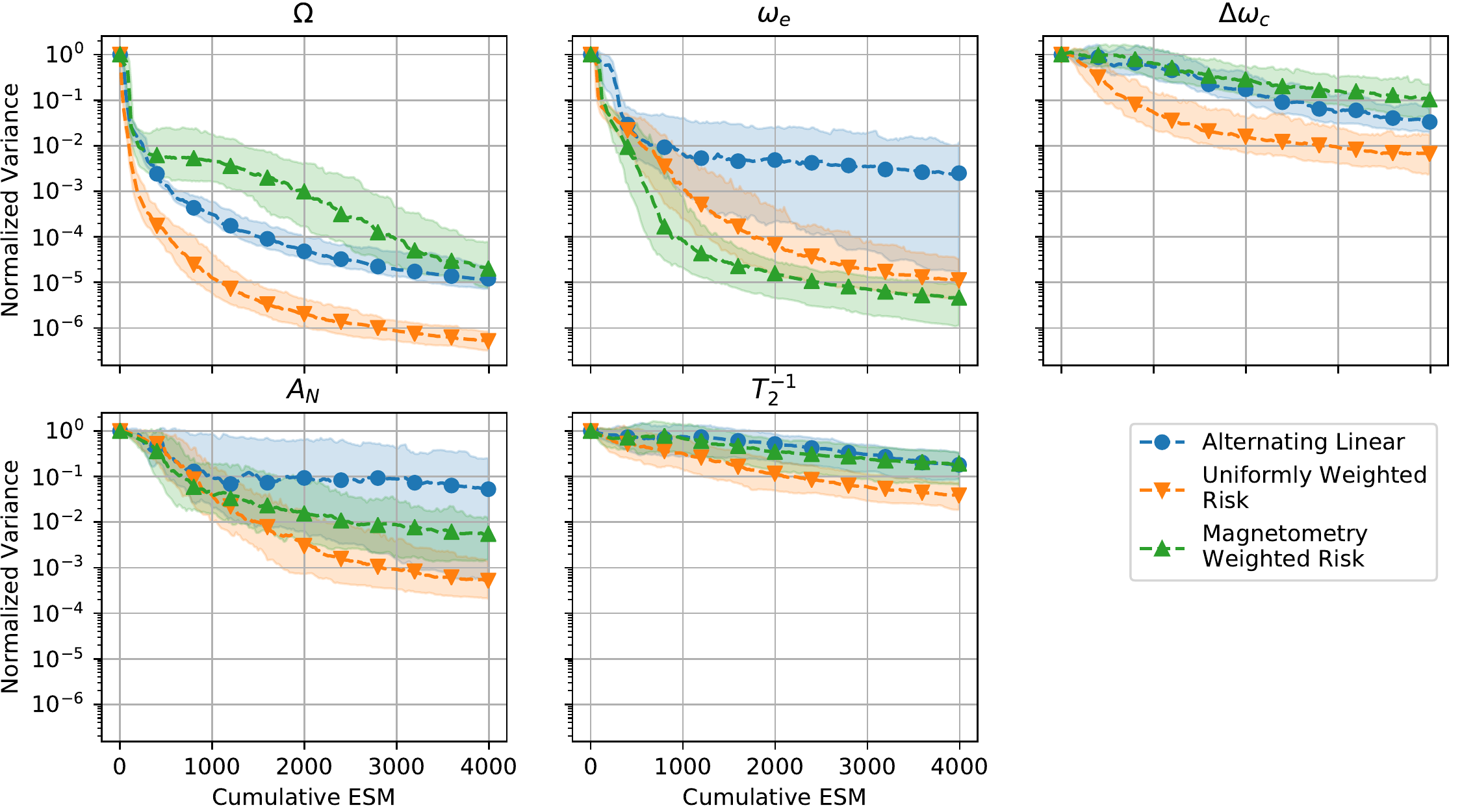}
    \caption{
        An extension of
        \subref{fig:heuristic-comparison}{(a-c)}
        that shows learning rates of all parameters relevant to
        the quantum dynamics of the system.
        }
    \label{fig:param-learning-rates}
\end{figure*}

\begin{figure*}
    \includegraphics[width=\textwidth]{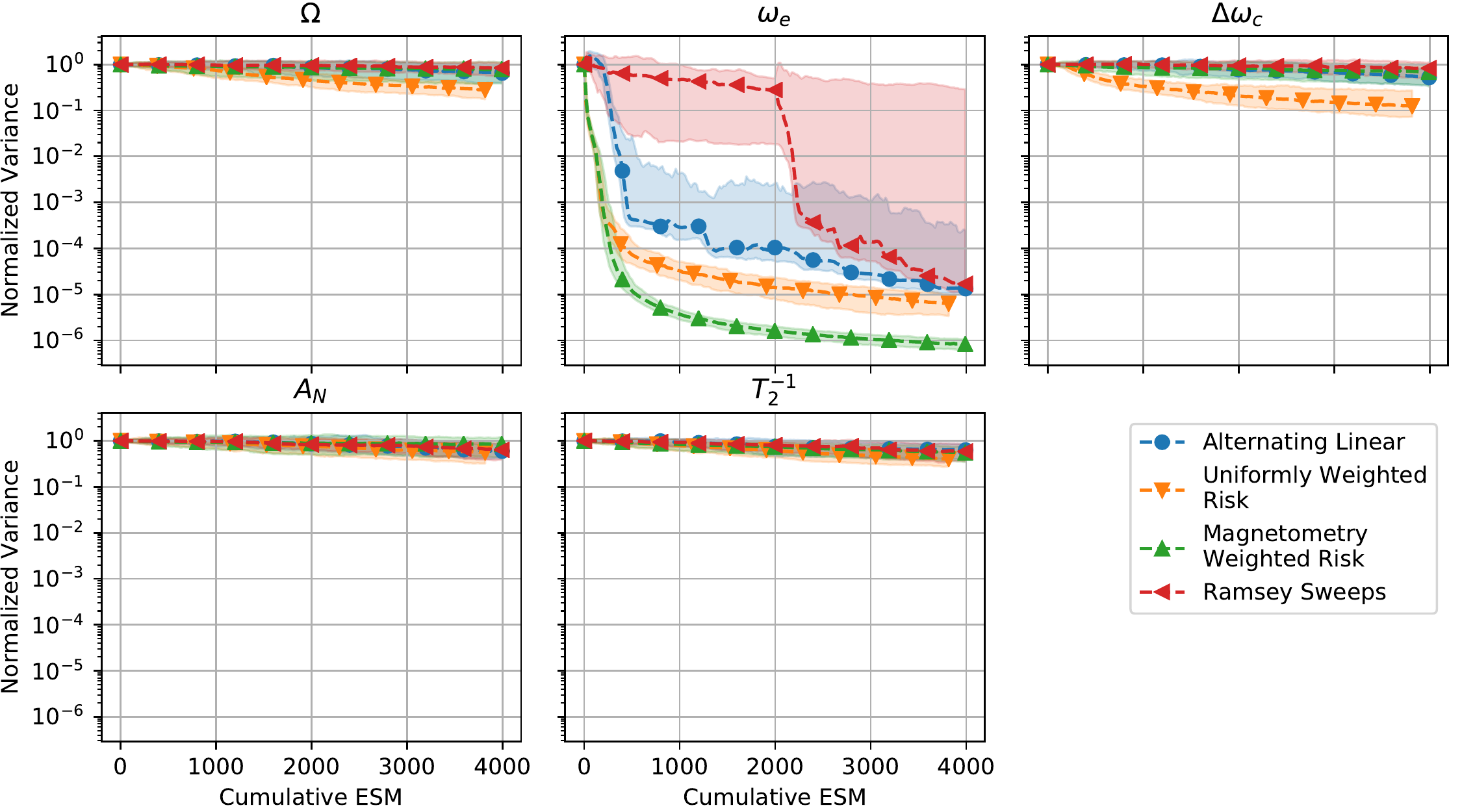}
    \caption{
        An extension of
        \subref{fig:heuristic-comparison}{(d-f)}
        that shows learning rates of all parameters relevant to
        the quantum dynamics of the system.
        }
    \label{fig:param-learning-rates-tight}
\end{figure*}

\begin{figure*}
    \includegraphics[width=\textwidth]{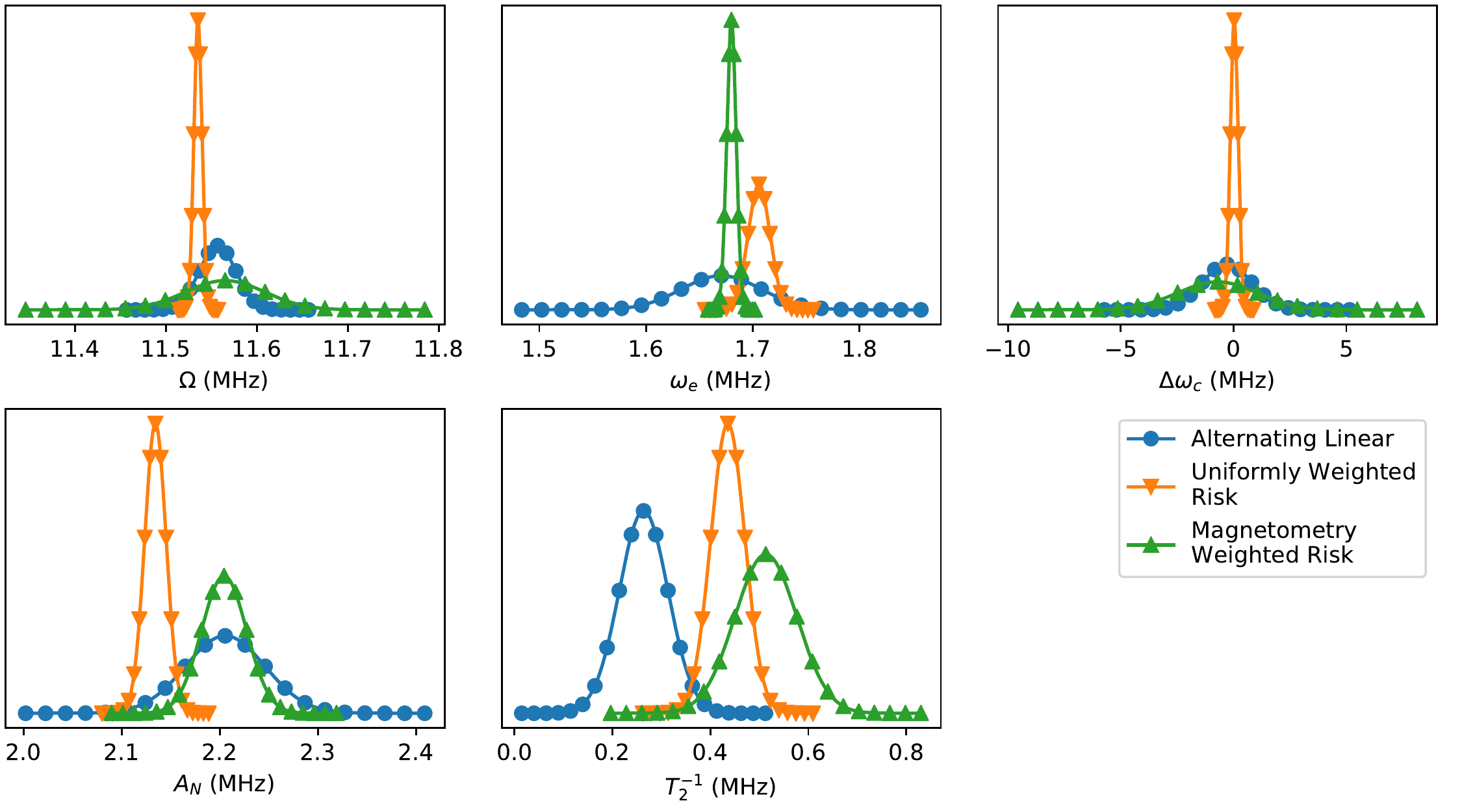}
    \caption{
        For each heuristic in
        \subref{fig:heuristic-comparison}{(a-c)},
        posterior marginal distributions are plotted for the first (of 100)
        trials on each parameter relevant to the quantum dynamics of the
        system.
        }
    \label{fig:param-posteriors}
\end{figure*}

\begin{figure*}
    \includegraphics[width=\textwidth]{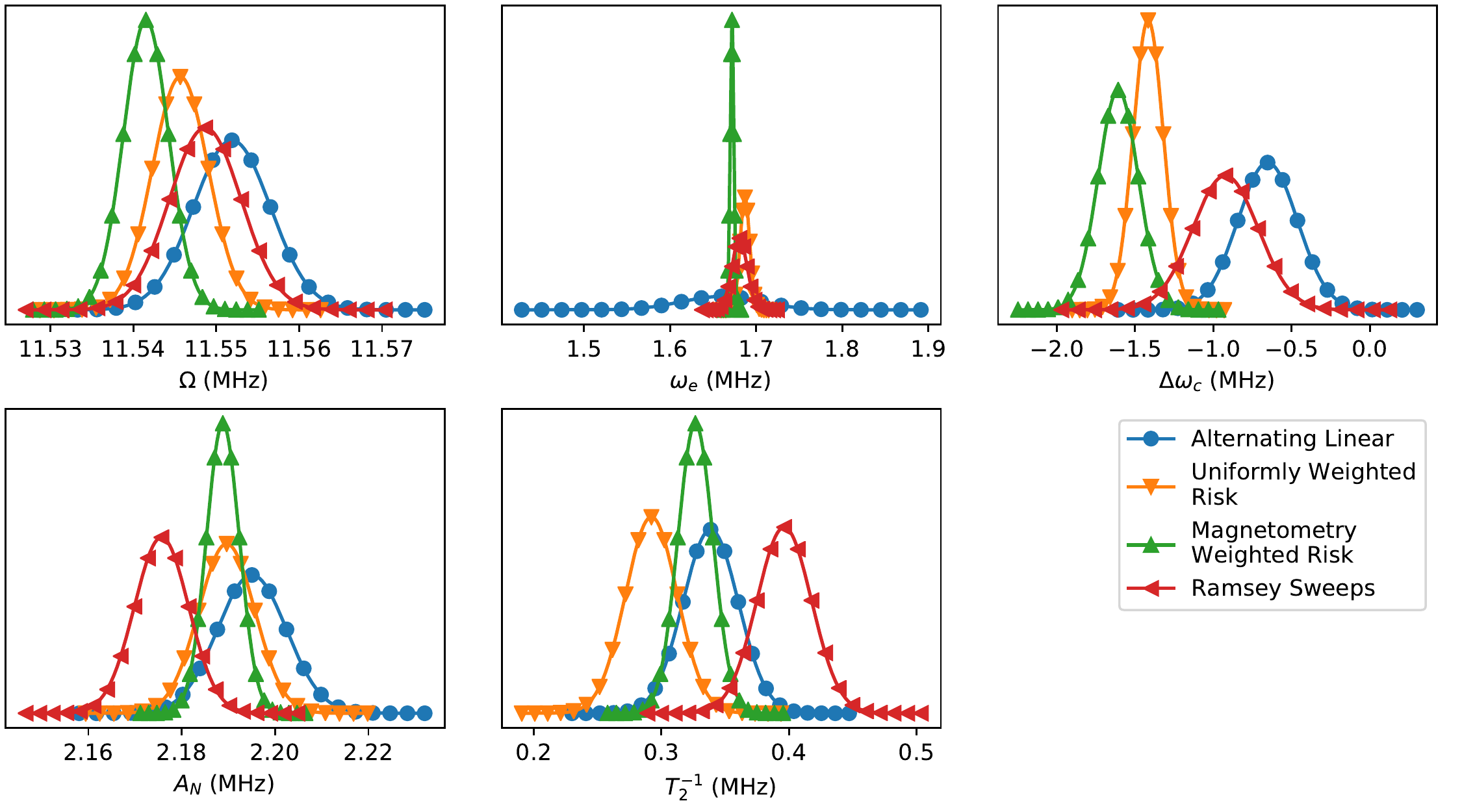}
    \caption{
        For each heuristic in
        \subref{fig:heuristic-comparison}{(d-f)},
        posterior marginal distributions are plotted for the first (of 100)
        trials on each parameter relevant to the quantum dynamics of the
        system.
        }
    \label{fig:param-posteriors-tight}
\end{figure*}

In \autoref{fig:heuristic-comparison}, only the learning rates of
$\omega_\text{e}$ are reported---in \autoref{fig:param-learning-rates} and
\autoref{fig:param-learning-rates-tight}, all learning rates are shown.
Posteriors are shown in \autoref{fig:param-posteriors} and
\autoref{fig:param-posteriors-tight}, where the first trial from each
heursitic is used as a representative.

%=============================================================================
\section{Effective Strong Measurements}
\label{apx:effective-strong-measurements}
%=============================================================================

Given a quantum state $\rho$,
information is accessed through the
Born's probability $p=\Tr(\ketbra{0}\rho)$.
In the hypothetical case of strong measurement, in the language
of statistics, we would be able to draw from
the Bernoulli distribution $\bernoullidist(p)$, or more generally, with
$n$ repeated preparations and strong measurements, from
the binomial distribution $\binomialdist(n,p)$.

Standard room temperature NV setups do not allow strong measurements.
Instead, access to the quantity $p$ is obstructed by three Poisson rates,
such that conditional on some values $0<\beta<\alpha$, we can
draw from the random variables
\begin{align}
    X|\alpha &\sim \poissondist(\alpha) \nonumber \\
    Y|\beta &\sim \poissondist(\beta) \nonumber \\
    Z|\alpha,\beta,p &\sim \poissondist(p\alpha + (1-p)\beta).
\end{align}
The quantities $\alpha$ and $\beta$ are known as the bright reference
and the dark reference, respectively.
They are defined as the
expected number of photons collected (and summed over $N$
repetitions of the experiment) using the initial NV states
$\ketbra{0}$ and $\ketbra{1}$, respectively\footnote{They are
more accurately defined in terms of the pseudo-pure states
that are actually created in the NV initialization procedure~\cite{hincks_statistical_2018}.}.

The information content about $p$ of this referenced Poisson model is not
immediately obvious,
and depends both on the magnitude of $\alpha+\beta$,
as well as the contrast between $\alpha$ and $\beta$.
This is different than the strong measurement case mentioned above,
where $n$ strong measurements has a clear intuitive and operational
interpretation.
The goal of this section is to reduce information about the references
$\alpha$ and $\beta$ into a single number
with the same interpretation as $n$.
This will allow one, for example,
to quantitatively compare two experimental setups or NVs and
decide which one is better at providing information about $p$.

It has been shown\cite{hincks_statistical_2018} that the
Fisher information matrix of this
referenced Poisson model is given by
\begin{align}
    J(p,\alpha,\beta) &=
        \begin{pmatrix}
            \frac{(\alpha -\beta )^2}{p (\alpha -\beta )+\beta } &
            \frac{p (\alpha -\beta )}{p (\alpha -\beta )+\beta } &
            \frac{\alpha }{\beta +\alpha  p-\beta  p}-1 \\
            \frac{p (\alpha -\beta )}{p (\alpha -\beta )+\beta } &
            \frac{p^2}{p \alpha -p \beta +\beta }+\frac{1}{\alpha } &
            -\frac{(p-1) p}{p (\alpha -\beta )+\beta } \\
            \frac{\alpha }{p \alpha -p \beta +\beta }-1 &
            -\frac{(p-1) p}{p (\alpha -\beta )+\beta } &
            \frac{p \alpha +(p-2) (p-1) \beta }{\beta  (p (\alpha -\beta )+\beta )} \\
        \end{pmatrix},
\end{align}
with inverse matrix
\begin{align}
    J(p,\alpha,\beta)^{-1}
        &=
        \begin{pmatrix}
            \frac{p (p+1) \alpha +(p-2) (p-1) \beta }{(\alpha -\beta )^2} &
            \frac{p \alpha }{\beta -\alpha } &
            \frac{(p-1) \beta }{\alpha -\beta } \\
            \frac{p \alpha }{\beta -\alpha } &
            \alpha  &
            0 \\
            \frac{(p-1) \beta }{\alpha -\beta } &
            0 &
            \beta
        \end{pmatrix}.
\end{align}

Using the Cramer-Rao bound, these matrices let us
estimate the information content of $p$ in the referenced
Poisson model.
Specifically, they give us an estimate in each of the following
extreme cases.
First, the $(p,p)$ element of $J^{-1}$,
$(J^{-1})_{p,p}=\frac{p (p+1) \alpha +(p-2) (p-1) \beta }{(\alpha -\beta )^2}$,
is a lower bound on the variance of any (unbiased) estimate of $p$ given that
a single measurement of the triple $(X,Y,Z)$ has been made,
with no prior information
about $p$, $\alpha$, or $\beta$ given.
Second, the inverse of the $(p,p)$ element of $J$,
$(J_{p,p})^{-1}=\frac{p (\alpha -\beta )+\beta}{(\alpha -\beta )^2 }$,
 is a lower bound on
the variance of any (unbiased) estimate of $p$ given that a single
measurement of $Z$ has been made, assuming perfect knowledge of
both $\alpha$ and $\beta$.

It will be useful for us to also be able to interpolate between these two
extremes, where some, but not all, prior information
about $\alpha$ and $\beta$ is available.
There are a few tacks that one might consider to achieve this, including
the Bayesian Cramer-Rao bound, or looking directly at the risk of some
estimator.
Instead, we choose a slightly ad-hoc method as it actually produces
a tractable calculation---statistics of the referenced Poisson model
usually involve a triple infinite sum, and many calculations are simply
not possible without numerics.
To this end, let $\sigma_\alpha^2$ and $\sigma_\beta^2$ represent
our prior variances of $\alpha$ and $\beta$, respectively, before
taking a measurement of $Z|\alpha,\beta,p$.
We can now ask the question: how many times, $M$, we must measure
$X|\alpha$ and $Y|\beta$ to produce these variances in the first
place?
We must allow $M$ to depend on $\alpha$ or $\beta$ in each case.
The distribution $\poissondist(M(\lambda)\lambda)$ has
Fisher information given by
$\frac{(M(\lambda)+\lambda M'(\lambda))^2}{\lambda M(\lambda}$.
Equating this to $1/\sigma^2$ and soliving the differential
equation at $M(0)=0$ gives $M=\lambda/4\sigma^2$.
Therefore consider the distribution
\begin{align}
    \poissondist\left(\frac{\alpha^2}{4\sigma_\alpha^2}\right)
        \times \poissondist\left(\frac{\beta^2}{4\sigma_\beta^2}\right)
        \times \poissondist\left(p\alpha+(1-p)\beta\right)
\end{align}
which effectively results in our desired information about $\alpha$
and $\beta$.
Solving for the $(p,p)$ element of the inverse Fisher information
matrix of this distribution results in the formula
\begin{align}
    K=\frac{
        \beta +p\left(\alpha-\beta+p \sigma_\alpha ^2
        +(p-2) \sigma_\beta ^2\right)+\sigma_\beta ^2}{(\alpha -\beta )^2}.
\end{align}
This formula correctly interpolates between the case of perfect prior
information, and prior information as collected by a single sample
of $(X,Y)|\alpha,\beta$, namely,
\begin{align}
    \lim_{\sigma_\alpha^2,\sigma_\beta^2\rightarrow 0} K
        &= (J_{p,p})^{-1} \quad\quad\text{and}\quad\quad
    \lim_{\sigma_\alpha^2\rightarrow \alpha, \sigma_\beta^2\rightarrow \beta} K
        = (J^{-1})_{p,p}.
\end{align}

The inverse Fisher information of the binomial model $\binomialdist(n,p)$
is given by $\frac{p(1-p)}{n}$, which when integrated uniformly over
$[0,1]$, produces $\frac{1}{6n}$.
Our definition for the number of effective strong measurements (ESM)
of a referenced Poisson model with parameters
$(\alpha,\beta,\sigma_\alpha,\sigma_\beta)$ is defined by
equating $\int_0^1 K\dd p=\frac{1}{6n}$ and solving for $n$,
which results in
\begin{align}
   \ESM = \frac{
            (\alpha-\beta)^2
        }{
            3(\alpha+\beta)+2\left(\sigma_\alpha^2+\sigma_\beta^2\right)
        }.
\end{align}
This shows, for example,
that having perfect information about $\alpha$ and $\beta$
before measuring $Z|\alpha,\beta,p$ is roughly equivalent---in terms
of information learned about p---to
$5/3\approx 1.67$ times more effective strong measurements
than the case where the triple $(X,Y,Z)|\alpha,\beta,p$ is
measured, but with no prior information.

\begin{figure}
    \includegraphics[width=\textwidth]{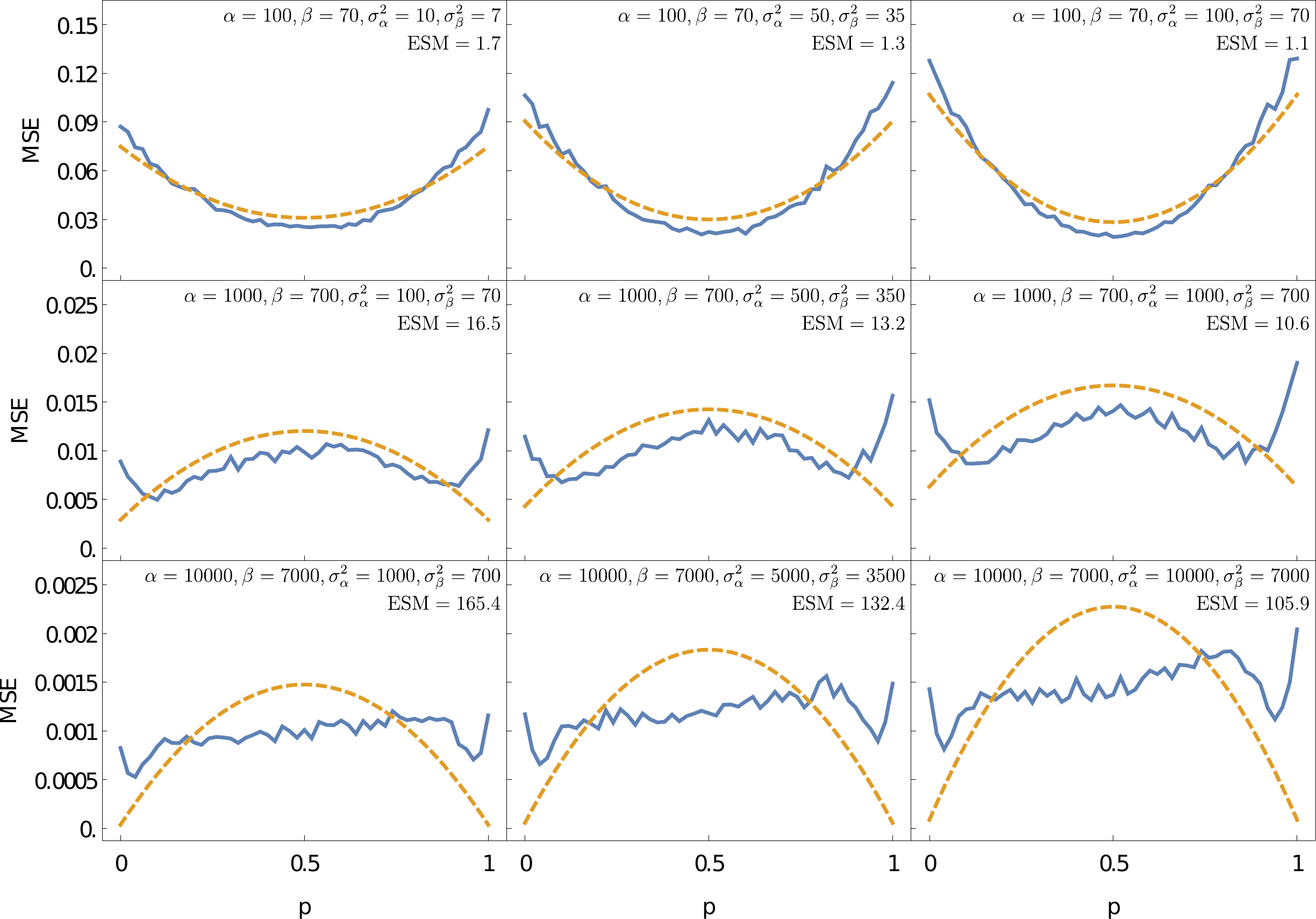}
    \caption{The mean-squared-error of the Bayes estimator is computed
    as a function of $p$
    for both the referenced Poisson model (blue, solid) and for a binomial model
    (orange, dashed) where $n=\ESM$.
    The prior distribution on $p$ is uniform.
    This is done in nine regimes, corresponding to the nine subplots of the figure.
    Each row uses a different magnitude of bright reference, $\alpha$,
    and each column uses a different amount of prior reference knowledge.
    The left column uses sub-Poisson error bars on $\alpha$ and $\beta$,
    and the right column uses regular Poisson error bars.}
    \label{fig:effective-strong-meas}
\end{figure}

Finally, in \autoref{fig:effective-strong-meas}, we use some numerics
to show that the $\ESM$
quantity accurately relates the mean-squared error of the Bayes estimator
for the referenced Poisson model and a binomial model with $n=\ESM$.

%=============================================================================
\section{Brute-force Numerical Evaluation of Bayes Risk}
\label{apx:brute-force-bayes-risk}
%=============================================================================

Evaluating the full Bayes risk for continuous outcome probability distributions
is not possible analytically apart from special cases such as linear models with
a normal likelihood function.
For finite outcome probability distributions the problem is more tractable,
however as the number of possible outcomes grows to be large, or even
infinite---such as the Poisson distributions considered within this paper---the
evaluation once again becomes intractable.

The difficulty of evaluation is a result of the expectation taken in
\autoref{eq:bayes-risk}.
For both infinite-discrete and continuous outcome probability distributions the
expectation is intractable, however for finite discrete distributions, the
expectation is a bounded discrete sum and straight-forward to evaluate
numerically.
We, therefore, aim to evaluate the Bayes risk by approximating the possible
outcomes with a finite, discrete set of outcomes---note that this technique may
also be used when the set of possible outcomes is finite but large enough to be
computationally intractable.
Typically, outcome domains are large with outcome probability mass
concentrated to a small portion of the outcome domain.
By fixing particle locations and sampling outcomes from these particles, we may
evaluate the risk for only the outcomes that ``matter'' within the regions of
outcome probability mass concentration.

We consider the case of evaluating the Bayes risk for the next experiment, $e$.
We assume throughout this discussion, that this hypothetical experiment $e$
was preceded by $n$ experiments $\eps_{1:n}$ with corresponding data
$\data_{1:n}$.
In several places, for brevity of notation, we will omit conditioning
on this prior information, for example, we have
$\Pr(\data|\eps)=\Pr(\data|\eps,\data_{1:n},\eps_{1:n})$.
We begin by re-approximating the particle filter prior distribution with a
uniformly weighted particle distribution by sampling $K'$ particles from the
prior $\pi_{n}$,
\begin{equation}
    \mps_j \sim \pi_{n}(\mps),
\end{equation}
which approximates the prior $\pi_{n}$ as
\begin{equation}
    \pi_n(\mps) \approx \frac{1}{K'}
        \sum \limits_{i}^{K'}\delta \left(\mps-\mps_i\right).
    \label{eq:approx_particle_filter}
\end{equation}
For each particle we now sample a datum from the likelihood function,
\begin{equation}
    \data^{(j)} \sim
        \Lhood \left(
            \mps_j;\data,\eps
        \right)
        \quad\quad \forall \ \mps_j \in \mps_{1:K'}.
\end{equation}
The set of sampled data is an approximation to the joint outcome, particle
distribution
\begin{equation}
    \Pr(\data,\mps|\eps) \approx
        \frac{1}{K'}\sum \limits_{i}^{K'} \delta \left(
            \data-\data^{(i)}\right)\delta\left(\mps-\mps_i
        \right).
    \label{eq:sampled-joint-distribution}
\end{equation}
The average utility---\autoref{eq:average-utility}---may be expanded in
conjunction with \autoref{eq:future-utility} as
\begin{align}
    U_n(\eps)
        &= \int\int \Pr(\data|\mps,\eps)\tilde\pi_{n,\data,\eps}(\mps)
            U_n(\mps,\data,\eps)\dd\mps \dd\data \nonumber \\
        &= \int\int \Pr(\data,\mps|\eps)
            U_n(\mps,\data,\eps)\dd\mps \dd\data.
\label{eq:average-utility-expanded}
\end{align}
The approximate particle, datum joint distribution,
\autoref{eq:sampled-joint-distribution} may be substituted into
\autoref{eq:average-utility-expanded} and the integrals are thus replaced by a
sum,
\begin{align}
    U_n(\eps) \approx
        \frac{1}{K'} \sum \limits_i^{K'} U_n(\mps,\data,\eps),
\end{align}
which is the average utility of the joint sampled particle, datum distribution.
When the utility is the negative mean-squared error the Bayes risk has the
approximate form
\begin{align}
    r_{n,Q}(\eps) \approx
        \frac{1}{K'}\sum \limits_i^{K'}
        \Tr Q
            (\mps_i - \hat\mps_{n,\data^{(i)},\eps})^\T
            (\mps_i - \hat\mps_{n,\data^{(i)},\eps}),
    \label{eq:bayes-risk-sampled}
\end{align}
where $\hat\mps_{n,\data^{(i)},\eps}$ is the posterior mean given the the
approximate prior \autoref{eq:approx_particle_filter},
\begin{align}
    \hat\mps_{n,\data^{(j)},\eps} =
        \sum \limits_i^{K'} \frac{\Lhood \left(
            \mps_i;d^{(j)},\eps
        \right)}{K'}\mps_i.
\end{align}
The evaluation of the Bayes risk requires on the order of $O(K'^2)$ likelihood
evaluations.
However, typically a large number of outcome samples will be required to
effectively sample the outcome domain of each particle and the total number of
outcome samples will roughly be $O(Kn_d)$, where $n_d$ is roughly the average
number of outcome datum desired per particle. This may be prohibitively large
when sampling is expensive.

Provided the outcome domain does not depend on the experiment---as is the case
for our experiments---we may perform maximum importance sampling (MIS) to sample
outcomes from an alternative distribution---chiefly the marginalized outcome
distribution $\Pr(\data|\eps)$---and properly re-weight the resultant utility
functions \cite{ueberhuber_numerical_1997}.
The sampled outcome distribution $\Pr(d|\eps)$ is obtained from the sampled
$\Pr(\data,\mps|\eps)$ by neglecting the associated model parameter,
\begin{equation}
    \Pr(\data|\eps) \approx
        \frac{1}{K'}\sum \limits_{i}^{K'} \delta
        \left(\data-\data^{(i)}\right).
    \label{eq:sampled-marginalized-distribution}
\end{equation}
The MIS utility is given as
\begin{align}
    U_n(\eps)
        &= \int\int \Pr(\data|\mps,\eps)\pi_n(\mps)
            \frac{\Pr(\data|\eps)}{\Pr(\data|\eps)}
            U_n(\mps,\data,\eps)
            \dd\mps \dd\data \nonumber\\
        &\approx \frac{1}{K'}\sum \limits_{i}^{K'}\sum \limits_{j}^{K}
            \frac{\Pr(\data^{(i)}|\mps_j,\eps)\omega_{n,j}}{
            \Pr(\data^{(i)}|\eps)}
            U_n(\mps_j,\data^{(i)},\eps)\nonumber \\
        &= \frac{1}{K'}\sum \limits_{i}^{K'}\sum \limits_{j}^{K} \omega_{n+1|
            \data^{(i)},j}U_n(\mps_j,\data^{(i)},\eps),
    \label{eq:mis-average-utility-expanded}
\end{align}
where the judicious choice of the sampling distribution has allowed the utility
to be written as the average of the posterior utility expectation  over the
marginalized outcome distribution.
For the case of the Bayes risk this may be further simplified to
\begin{align}
    r_{n,Q}(\eps) &\approx
        \frac{1}{K'}\sum \limits_i^{K'}  \Tr\left[ Q  \Cov_{\tilde{\pi}}
        [\mps|\data^{(i)},\eps]\right] \nonumber\\
        &=\frac{1}{K'}\sum \limits_i^{K'}
            \Tr\left[ Q(\widehat{x^Tx}_{i}-\hat{x}_i^T\hat{x}_i)\right],
    \label{eq:mis-bayes-risk-sampled}
\end{align}
where $\widehat{x^Tx}_i=\sum
\limits_{j}^K\omega_{n+1|\data^{(i)},j}x_j^Tx_j$, and $\hat{x}_i=\sum
\limits_{j}^K\omega_{n+1|\data^{(i)},j}x_j$.

In general the the initial prior distribution may be down-sampled to some number
of particles $K$, such that we now have two parameters that may be tuned, the
number of outcome samples $K'$, and the number of model parameter particles $K$.
With the MIS Bayes risk, the number of likelihood function calls is now
$O(KK')$, with only $O(K')$ outcome samples required.
We utilize the MIS Bayes risk for experiment design within this paper.

\begin{figure*}
    \includegraphics[width=\textwidth]{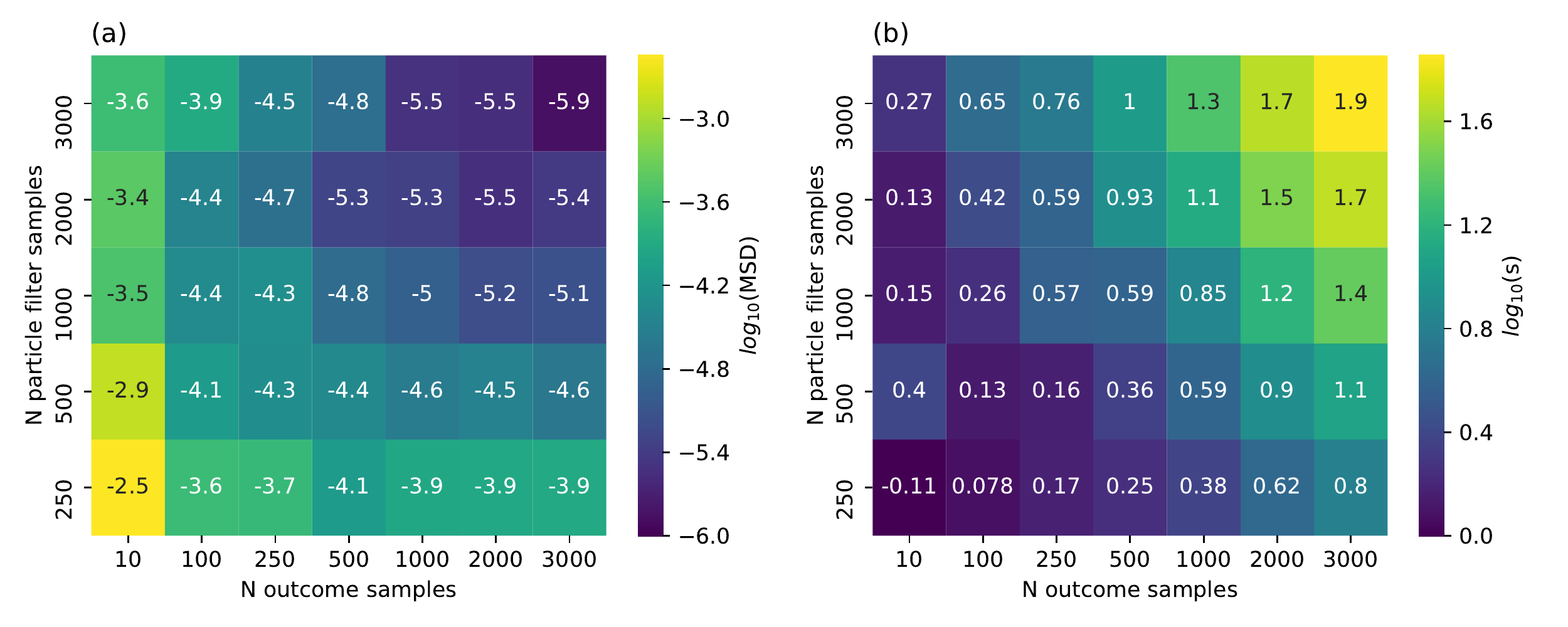}
    \caption{Comparison of various outcomes and particle sampling accuracies and
        times when evaluating the MIS Bayes risk.
        The prior distribution over model parameters is given by
        \autoref{eq:wide-prior}, and the experiments which the Bayes' risk is
        computed for is the uniformly weighted experiment design risk heuristic
        found in \autoref{tab:heuristics}. (a) Log mean squared difference for
        all experiments computed with respect to a 4000 outcome, 4000 particle
        reference evaluation. (b) Log evaluation time(s) of Bayes risk over all
        experiments for a given number of outcome and particle samples.
        These calculations were done on an i9-7980XE CPU.}
    \label{fig:computational-cost-heatmap}
\end{figure*}

In practice a trade-off between accuracy and computational cost/time is necessary
when selecting the number of outcomes and particle samples, $K'$ and $K$
respectively for the evaluation of the MIS Bayes risk.
A comparison of various sampling numbers is displayed in the heatmaps of
\autoref{fig:computational-cost-heatmap}, which were evaluated with the wide
prior of \autoref{eq:wide-prior} and experiments for the uniformly weighted
Bayes risk experiment design heuristic given in \autoref{tab:heuristics}.
The aquistion of 4000 \ESM\ takes roughly 10 seconds, as the full particle
filter update of 30000 particles takes roughly 2 seconds, there are 8 seconds
remaining in which to compute the Bayes risk and select the optimal experiment.
We use $K'=512$ outcome samples and $K=1024$ particle filter samples, as this
strikes a balance between accuracy while keeping the evaluation time
below our threshold on our computational hardware.
As this problem is massively parallel, if desired it is simple to use additional
computational resources to refine to the evaluation accuracy.

\end{document}